\documentclass[aps,pra,twocolumn,reprint,floatfix, superscriptaddress,longbibliography ]{revtex4-2}

\usepackage{graphicx}
\usepackage{dcolumn}
\usepackage{bm}

\usepackage{amsmath} 
\usepackage{amssymb} 
\usepackage{physics}
\usepackage{xcolor}
\usepackage{subfiles}

\begin{document}

\preprint{APS/123-QED}

\title{Photon Squeezing in Photonic Time Crystals}

\author{Jaime Echave-Sustaeta}
\email{jaime.echave-sustaeta@uam.es}
\affiliation{Departamento de F\'{i}sica Te\'orica de la Materia Condensada, Universidad Aut\'onoma de Madrid, E-28049 Madrid, Spain}
\affiliation{Condensed Matter Physics Center (IFIMAC), Universidad Aut\'onoma de Madrid, E-28049 Madrid, Spain}

\author{Francisco J. García-Vidal}
\affiliation{Departamento de F\'{i}sica Te\'orica de la Materia Condensada, Universidad Aut\'onoma de Madrid, E-28049 Madrid, Spain}
\affiliation{Condensed Matter Physics Center (IFIMAC), Universidad Aut\'onoma de Madrid, E-28049 Madrid, Spain}
\author{P.A. Huidobro}
\email{p.arroyo-huidobro@uam.es}
\affiliation{Departamento de F\'{i}sica Te\'orica de la Materia Condensada, Universidad Aut\'onoma de Madrid, E-28049 Madrid, Spain}
\affiliation{Condensed Matter Physics Center (IFIMAC), Universidad Aut\'onoma de Madrid, E-28049 Madrid, Spain}
\affiliation{Instituto de Telecomunica\c c\~oes, Instituto Superior Tecnico-University of Lisbon, Avenida Rovisco Pais 1, Lisboa, 1049-001 Portugal}

\date{\today}

\begin{abstract}
Time-varying media offer a platform to realize novel and exotic wave effects, including photonic time crystals characterized by momentum band gaps with exponential wave amplification. Here we focus on the quantum electrodynamical properties of time-varying media, in particular vacuum amplification and squeezing. For that purpose, we present a theory of photon pair generation in photonic time crystals that unveils the link between the classical and quantum electrodynamical properties of these systems, that is, a direct relation between reflectivity and pair generation through the squeezing parameter. By working within an Hermitian framework, we are able to characterize quantum pair generation processes in photonic time crystals, showing how momentum bandgaps result in a non-resonant exponential enhancement of dynamical Casimir processes.   
\end{abstract}

\maketitle


Time-varying media open new doors to controlling the propagation of electromagnetic waves \cite{galiffi2022photonics,pacheco2022time,engheta2023four,boltasseva2024photonic}. When the optical parameters of a material are modulated in time, a myriad of new physical phenomena with no counterpart in spatially structured systems is unravelled. These include temporal refraction and reflection \cite{moussa2023observation}, frequency conversion \cite{zhou2020broadband, solis2021time,pang2021adiabatic,liu2021photon}, temporal diffraction \cite{tirole2023double}, a Fresnel light drag based on synthetic motion \cite{huidobro2019fresnel,prudencio2023replicating}, spatio-temporal metasurfaces \cite{wang2020theory}, spontaneous emission from stationary sources \cite{li2023stationary}, temporal coherent wave control \cite{chamanara2019simultaneous,galiffi2023broadband}, axion-like non-reciprocal couplings \cite{prudencio2023synthetic} or the possibility of studying black hole analogues \cite{pendry2022photon,bahrami2023electrodynamics,horsley2023quantum}, all of which stem from the time reversal symmetry breaking taking place in these system.
\par

The simplest example of a time-varying medium is a temporal interface: an instantaneous, homogeneous and isotropic change of the optical properties of a material \cite{morgenthaler1958velocity,felsen1970wave,fante1971transmission,Ramaccia:20,ramaccia2021temporal,mirmoosa2024time}. Such a system conserves momentum, but not frequency, and vertical transitions between different light-cones take place \cite{martinez2018parametric,apffel2022frequency}. Lack of energy conservation allows the incident (forward) wave to change frequency and be amplified, but in order for momentum to be the same before and after the switch, a backward wave emerges, also at a new frequency. These waves are manifestations of time refraction and reflection phenomena \cite{bacot2016time,fink2017time}, which have been observed for electromagnetic waves with transmission lines \cite{moussa2023observation}. Stacking more than one of these interfaces allows for interference between backward waves to occur, with some modes even becoming transparent to the modulation in a temporal analogue of an antireflection coating \cite{pacheco2020antireflection}.
\par
Periodic temporal modulations of the optical parameters result in a photonic time crystal (PTC), which displays band structures with momentum bandgaps where frequency is complex-valued \cite{martinez2018parametric,boltasseva2024photonic}. PTCs have been experimentally realized in the microwave regime \cite{reyes2015observation,wang2023metasurface}. For their realization at higher frequencies,  low Drude weight semiconductors such as indium tin oxide (ITO) are promising candidates, since they enable an ultrafast and unprecedentedly strong modulations of the refractive index \cite{liberal2017near,boyd2017beyond,reshef2019nonlinear,bohn2021all,vezzoli2018optical,tirole2022saturable,lustig2023time}. 

\par
On the other hand, time varying media also offer a very rich platform from the point of view quantum electrodynamical effects  
\cite{mendoncca2000quantum,mendoncca2003temporal,lyubarov2022amplified,vazquez2023shaping,horsley2023quantum,liberal2023quantum,ganfornina2023quantum}. Through the interaction between quantum fluctuations and the dynamic properties of time-varying media, these systems allow to amplify the quantum vacuum \cite{nation2012colloquium}. In particular, pairs of photons can be spontaneously created from the vacuum in a squeezed state at a temporal interface \cite{mendoncca2003temporal}, through the dynamic Casimir effect \cite{wilson2011observation,nation2012colloquium,lahteenmaki2013dynamical,dodonov2020fifty}. Interestingly, time varying media offer great control over vacuum amplification processes. For instance, anisotropic temporal boundaries provide angular control over vacuum amplification \cite{vazquez2023shaping}, and quantum antireflection temporal coatings induce a frequency shift of the quantum state while preserving photon statistics \cite{liberal2023quantum}. Furthermore, synthetically moving gratings also result in radiation from the quantum vacuum, in an analogue to Hawking radiation \cite{horsley2023quantum}. However, a clear connection between quantum vacuum processes and PTCs has not yet been drawn, in part due to difficulties in formulating a quantum theory of the momentum band gaps that emerge in this open system. 

\par

In this Letter, we present a theory of photon pair creation in time-varying media in the PTC regime. For this purpose, we first consider the classical electrodynamics of PTCs and show that the physical properties that characterize this regime emerge already when just a few temporal boundaries are considered. Next, we introduce field quantization through a transfer matrix approach. This allows us to describe photon pair generation within an Hermitian theory, even within the momentum bandgaps. In doing so, we provide a direct link between the properties of classical light in PTCs, such as the reflectivity, and quantum amplification effects. This way, we describe photon pair generation and squeezing in PTCs, showing how momentum bandgaps result in a non-resonant exponential enhancement of dynamical Casimir processes. 

\par
\begin{figure*}
    \centering
    \vspace*{0cm}\hspace*{-0cm}\includegraphics[scale = 0.55]{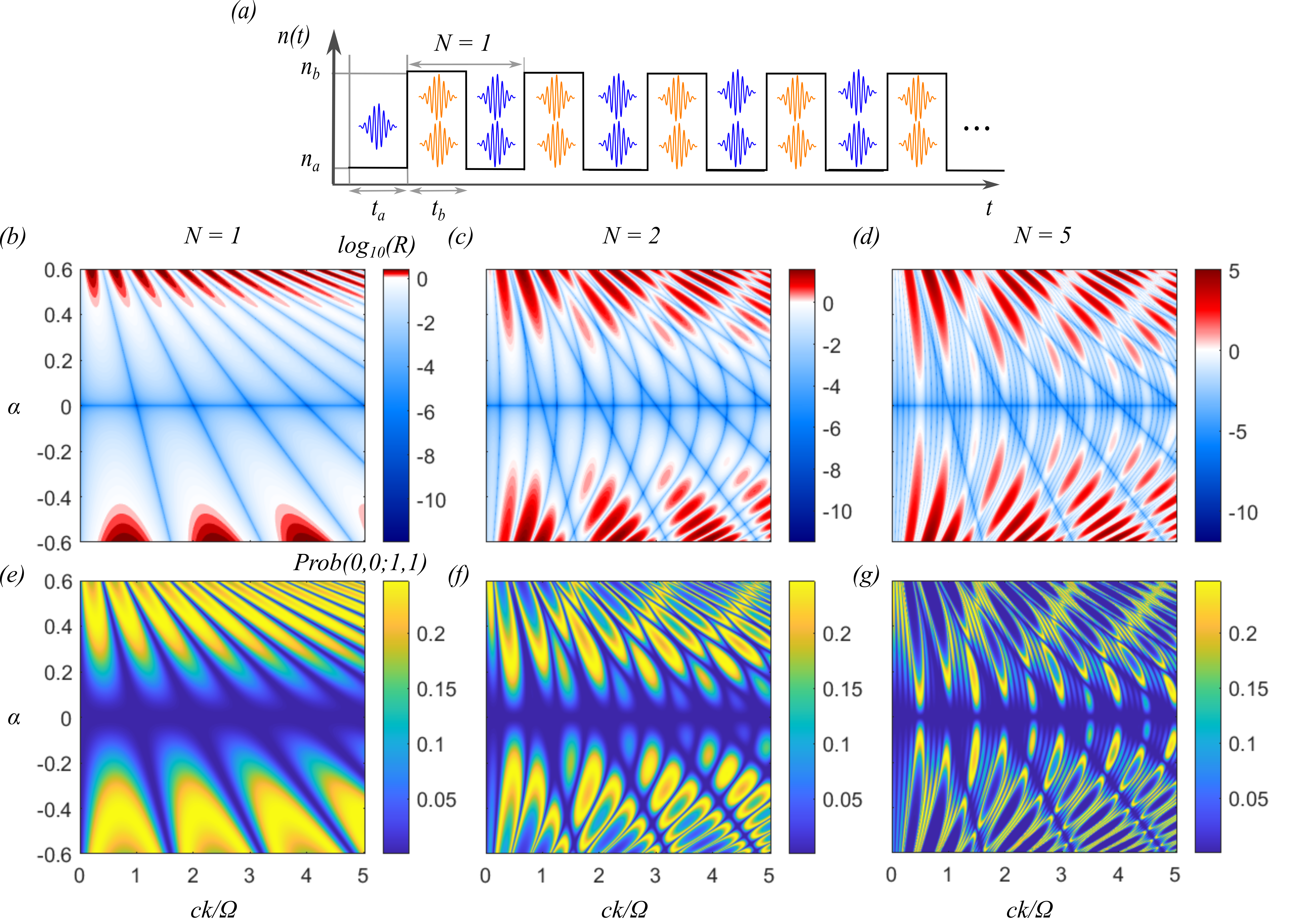}
    \caption{(a) Sketch of the PTC: the basic unit of the transfer matrix is a $n_b$ slab followed by a $n_a$ one. (b-d) Reflectivity, $\mathcal{R}$ (logarithmic scale) for $N = 1 \text{ (b)}, N = 2\text{ (c) and } N = 5 \text{ (d)}$ periods of the time modulation. (e-g) Probability for single photon pair creation from the vacuum, again for $N = 1 \text{ (e)}, N = 2\text{ (f) and } N = 5 \text{ (g)}$ periods.}
    \label{fig:wide_fig_1}
\end{figure*}
\par
\textit{Transfer matrix and time evolution}. 
Let us consider a spatially homogeneous and isotropic medium whose permittivity $\varepsilon(t)$, and thus refractive index $n(t)$, is periodically modulated in time, in a series of instantaneous temporal interfaces. 
The refractive index alternatively takes two values, $n_a$ and $n_b$; we shall name each of these regions with constant refractive index a \textit{temporal slab} or \textit{slab} for abbreviation. Each slab with $n(t) = n_a$ lasts for a time $t_a$, while those with $n(t)=n_b$ do so for a time $t_b$, with $T = t_a + t_b$ being the period of the PTC. Such a system is depicted in Fig. 1(a). Since the medium is homogeneous and isotropic, both the wavevector $\boldsymbol{k}$ and the polarization $\lambda$ are conserved quantities; however, as commented above, the frequency is not, and vertical transitions take place between the light cones of adjacent slabs \cite{martinez2018parametric}. Applying temporal boundary conditions to Maxwell equations (see Supplemental Material for details \cite{suppmat}), we build up a transfer matrix and use it to connect the field amplitudes of different $n(t) = n_a$ temporal slabs \cite{li2023stationary}, as
\begin{equation}
\begin{pmatrix}
    (B_{-\boldsymbol{k}\sigma}^{(N)})^* \\ \\B_{\boldsymbol{k}\sigma}^{(N)}
\end{pmatrix} = \boldsymbol{T}^N\cdot \begin{pmatrix}
    (B_{-\boldsymbol{k}\sigma}^{(0)})^* \\ \\ B_{\boldsymbol{k}\sigma}^{(0)}
\end{pmatrix}.
\end{equation}
Here $B^{(N)}_{\boldsymbol{k}\sigma}$ is the amplitude of the magnetic induction flux within the $N$-th $n_a$ type slab, with wavevector $\boldsymbol{k}$ and polarization $\sigma$, and  $\boldsymbol{T}$ is the transfer matrix. The $\boldsymbol{T}-$matrix couples forward ($\boldsymbol{k}$) and backward modes ($-\boldsymbol{k}$) at the 0-th and $N$-th slabs, and guaranties that momentum is conserved. Since the $\boldsymbol{T}-$matrix does not couple modes with different polarizations, we shall omit the $\sigma$ subscript from now on.
\par

From the eigenvalues ($\lambda_{\pm}$) of the $\boldsymbol{T}-$matrix, it is possible to define a Floquet frequency $\omega_F$ through $\lambda_{\pm} = \exp(\mp i\omega_FT)$. For some values of the wavevector $k$, the Floquet frequency $\omega_F$ becomes complex-valued and amplification of electromagnetic waves takes place. These $k$ intervals with complex $\omega_F$ define the momentum band gaps of the PTC, which can be populated due to energy not being conserved in these systems \cite{khurgin2023photonic}. This way, the $\boldsymbol{T}-$matrix enables us to study the emergence of the PTC regime as more layers are added to the system. 

As we are dealing with real solutions of Maxwell equations and their temporal boundary conditions, the following properties of the transfer matrix can be inferred: $(\boldsymbol{T}^N)_{11} = \big((\boldsymbol{T}^N)_{22}\big)^*$ and $(\boldsymbol{T}^N)_{12} = \big((\boldsymbol{T}^N)_{21}\big)^*$, as well as $\det(\boldsymbol{T}) = 1$. These properties enable us to write 
\begin{equation}
    (\boldsymbol{T}^N)_{11} = \cosh(r)e^{i\theta_1},\text{  } (\boldsymbol{T}^N)_{12} = \sinh(r)e^{i\theta_2},
    \label{eq:t_matrix_elements}
\end{equation}
where $\theta_1$ and $\theta_2$ are respectively the phases of forward and backscattered waves, and $r$ measures the strength of forward amplification and backscattering. Furthermore, the classical transmitivity and reflectivity of the PTC are given by $\mathcal{T} = |(\boldsymbol{T}^N)_{11}|^2 = \cosh^2(r)$ and $\mathcal{R} = |(\boldsymbol{T}^N)_{12}|^2 = \sinh^2(r)$, respectively. As a consequence of momentum conservation, $\mathcal{T}-\mathcal{R} = 1$. From this constraint we see $\mathcal{T}\geq1$, meaning the temporal modulation can only enhance (or be transparent to) an incoming signal, but never suppress it, and always at the expense of some backscattering.
\par
Both the transmitivity and reflectivity depend on the wavenumber $k$, the two values taken by the refractive index $n_a$ and $n_b$, the duration of each slab $t_a$ and $t_b$, and lastly, the number of periods $N$ of the modulation. For the sake of simplicity, we assume $t_a = t_b$, and $n_a = 1 + \alpha$ and $n_b = 1 - \alpha$, with $-1<\alpha<1$ measuring the strength of the modulation. In Figs. \ref{fig:wide_fig_1}$(b)-(d)$ we show the reflectivity for $N = 1, 2 \text{ and } 5$ slabs, as a function of both $\alpha$ and $ck/\Omega$, where $c$ is the speed of light in the unmodulated medium ($\alpha=0$) and $\Omega = 2\pi/T$ is the modulation frequency. The red regions correspond to a reflectivity larger than unity and hence, to a strong backscattering, while the blue ones exhibit a comparatively weaker temporal reflection. For $N=1$, see panel (b), high backscattering regions appear for large values of $|\alpha|$. However, these red regions progressively tend towards the $\alpha = 0$ horizontal line (in which $\mathcal{R}=0$) as the number of periods of the PTC increases [see panels (c,d)]. This means that strong modulations are not needed in order to have amplification of waves if the periodic modulation is sustained for a sufficiently long time.
\par
On the other hand, for $N=1$ (b), we see dark blue lines with negative slope in which the reflectivity vanishes: these are the transparency lines of the temporal slab, which satisfy $(\boldsymbol{T})_{12}=0$, the antireflection temporal coating condition \cite{pacheco2020antireflection,liberal2023quantum,galiffi2023broadband}, and come from destructive interference between waves backscattered at $t = t_a$ and $t = t_a + T$. As the number of periods of the PTC increases, new transparency lines start emerging, as seen in panels (c) and (d). These are the zeros of $(\boldsymbol{T}^N)_{12} = (\sin(N\omega_FT)/\sin(\omega_FT))(\boldsymbol{T})_{12}$, and, as in the $N = 1$ case, come from the destructive interference of waves backscattered at different time interfaces. As seen in both panels, the more slabs are added to the PTC, the more transparency lines accumulate, as can also be inferred from the $N$-dependence of the formula for $(\boldsymbol{T}^N)_{12}$. 
If a mode becomes transparent to the modulation at a given number of periods, it will remain so in the future.
\par
Thus, from our approach we can draw conclusions about the emergence of the PTC regime with just a few temporal slabs, $N\ge2$. While some broad range of values of momentum entail an exponential amplification of modes [red areas in Figs. \ref{fig:wide_fig_1}(c-d)], transparency lines progressively accumulate with increasing $N$ (dark blue areas in the plots). The first phenomenology correspond to the momentum band gaps, and the second to the bands, a correspondence that we will see more clearly below. 
\par
\textit{Photon pair creation}. Next, we quantise the field within each temporal slab \cite{suppmat}. Employing canonical quantization, the photon operators of different slabs can be shown to be connected through the same transfer matrix as the classical fields. This connection results in a Bogoliubov transformation of the operators along the PTC, \cite{mendoncca2003temporal,liberal2023quantum,vazquez2023shaping},
\begin{eqnarray}
    (\Hat{a}_{-\boldsymbol{k}}^{(N)})^\dagger = \cosh(r)e^{i\theta_1}(\Hat{a}_{-\boldsymbol{k}}^{(0)})^\dagger + \sinh(r)e^{i\theta_2}\hat{a}^{(0)}_{\boldsymbol{k}},
    \label{eq:bogoliubov_1} \\
    \hat{a}^{(N)}_{\boldsymbol{k}} = \cosh(r)e^{-i\theta_1}\hat{a}^{(0)}_{\boldsymbol{k}} + \sinh(r)e^{-i\theta_2}(\Hat{a}_{-\boldsymbol{k}}^{(0)})^\dagger,
    \label{eq:bogoliubov_2}
\end{eqnarray}
where $\hat{a}^{(N)}_{\boldsymbol{k}}$ annihilates a photon with wavevector $\boldsymbol{k}$ for $t\in[NT,NT+t_a)$, and correspondingly for the creation operators. Additionally, $r$ and $\theta_{1,2}$ come from the classical transfer matrix, see Eq. (2). On the other hand, and as can be seen from Eqs. (\ref{eq:bogoliubov_1})-(\ref{eq:bogoliubov_2}), such transformation between forward and backward photon operators implements a squeezing operation \cite{barnett2002methods,milburn2012quantum,gerry2023introductory}. Thus, by defining the complex squeezing parameter 
\begin{equation}
    \zeta = r\exp(i\varphi),
\end{equation}
where we recall that $r$ gives the classical reflectivity, $\mathcal{R}=\sinh^2(r)$, and where $\varphi = \theta_1-\theta_2$, we can introduce the following two-mode squeezing operator
\begin{equation}
    \hat{S}(\zeta) = \exp(\zeta\Hat{a}^{(0)}_{\boldsymbol{k}}\Hat{a}^{(0)}_{-\boldsymbol{k}}-\zeta^*(\Hat{a}_{\boldsymbol{k}}^{(0)})^\dagger(\Hat{a}_{-\boldsymbol{k}}^{(0)})^\dagger),
\end{equation}
which acts as the unitary time-evolution operator in this theory \cite{suppmat}. Thus, we see how each time interface results in the generation of photon pairs, with forward and backward photons being correlated in a squeezed state owing to momentum conservation \cite{mendoncca2003temporal,liberal2023quantum,vazquez2023shaping}.
\par
Critically, this theoretical approach allows us to describe the momentum bandgaps of PTCs and their corresponding complex Floquet frequencies within a unitary and hence probability-conserving framework. Photon amplification within the PTC band gaps can be seen as a cascade of creation processes of squeezed photon pairs at subsequent temporal interfaces. This is in contrast to previous quantum mechanical descriptions of PTCs, which have so far neglected its bandgaps. While a Floquet expansion of the fields would result in a non-Hermitian theory \cite{lyubarov2022amplified,park2022comment,ashida2020non}, our transfer matrix approach does not rely on the existence of a complete basis and avoids this complication. 
\par
With our framework, we can compute photon transition probabilities. These can be obtained from the matrix elements of the squeezing operator in the number state basis, 
\begin{equation}
    \begin{split}
        &\bra{n_{\boldsymbol{k}}',m_{-\boldsymbol{k}}'}\hat{S}(\zeta)\ket{n_{\boldsymbol{k}},m_{-\boldsymbol{k}}} = \frac{\left(-e^{i\varphi}\tanh(r)\right)^{n'-n}}{(\cosh(r))^{n+m+1}}\\&\sum_{l = \max(0,n-n')}^{\min(n,m)}C^l_{n,m;n',m'}(-\sinh^2(r))^l\delta_{n'-n,m'-m},
    \end{split}
    \label{eq:two_mode_squeezing}
\end{equation}
where
\begin{equation}
        C^l_{n,m;n',m'} = \frac{\sqrt{n'!m'!n!m!}}{l!(l+n'-n)!(n-l)!(m-l)!},
    \label{eq:two_mode_squeezing_coeffs}
\end{equation}
and where the Kronecker delta in Eq. \ref{eq:two_mode_squeezing} ensures momentum conservation. The probability for the transition $\ket{n_{\boldsymbol{k}},m_{-\boldsymbol{k}}}\longrightarrow\ket{n'_{\boldsymbol{k}},m'_{-\boldsymbol{k}}}$ is then given by $\text{Prob}(m,n;n',m') = |\bra{n_{\boldsymbol{k}}',m_{-\boldsymbol{k}}'}\hat{S}(\zeta)\ket{n_{\boldsymbol{k}},m_{-\boldsymbol{k}}}|^2$.
\par
In Figs. \ref{fig:wide_fig_1}$(e)-(g)$ we plot the probability of creating a single photon pair starting from the vacuum state, i.e.,  $|\bra{1_{\boldsymbol{k}},1_{-\boldsymbol{k}}}\hat{S}(\zeta)\ket{0_{\boldsymbol{k}},0_{-\boldsymbol{k}}}|^2$, for the same values of $N$ as considered in panels (c-d). For the case of a single temporal slab, $N=1$, a correspondence between the classical reflectivity of the PTC, $\mathcal{R} = \sinh^2(r)$, and the quantum transition probabilities between number states can be infered by comparing panels (b) and (e). As may be expected intuitively, we find a connection between strong backscattering and high photon pair creation probabilities, while weak backscattering implies low photon pair creation probabilities. 
However, for $N=2$ we see that this correspondence between the classical and quantum quantities starts to disappear. In particular, the regions of largest reflectivity and sufficient large $|\alpha|$ in panel (c), do not correspond to regions of largest photon pair creation probability in panel (f), but to regions of very low one. Results for larger number of slabs ($N=5$), show that this effect is even more pronounced [compare panels (d) and (g)], and the (yellow) pockets of high pair creation probability become smaller and migrate towards lower values of $|\alpha|$. 
Since our theory is unitary, there cannot be a probability leakage to a surrounding environment; as we will show below, what is happening is that the probability, which is conserved, is migrating towards higher order processes $\ket{0_{\boldsymbol{k}},0_{-\boldsymbol{k}}}\longrightarrow\ket{m_{\boldsymbol{k}},m_{-\boldsymbol{k}}}$, for $m>1$, thus reducing the probability of creating a single pair ($m=1$). Interestingly, photon pair creation probability in PTCs is quite broadband in the photon free-space frequency $ck/\Omega$. This can be understood from the fact that in time-varying media energy is not conserved and that, contrary to the optical parametric amplifier \cite{boyd2008nonlinear,drummond2014quantum}, there is no frequency matching to be made between the pump and the amplified waves.
\par
\begin{figure}
    \centering
    \hspace*{-0.5cm}
    \includegraphics[scale = 0.55]{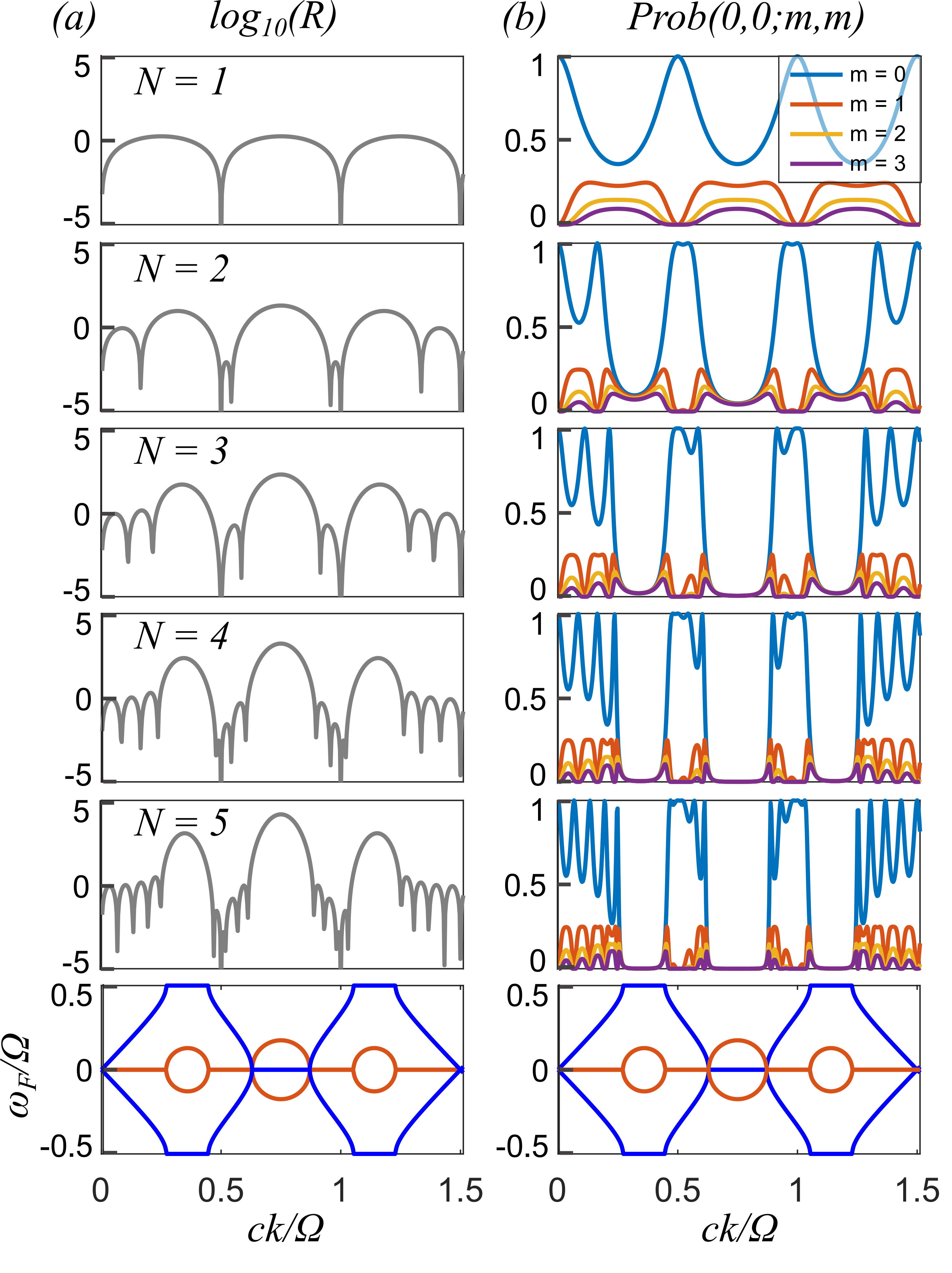}
    \caption{Evolution of PTC classical (a) and quantum (b) properties with number of temporal slabs, for a fixed modulation strength ($\alpha = 0.5$). 
    (a) Reflectivity, $\mathcal{R}$, is shown in the top 5 panels ($N=1,2,3,4\text{ and }5$, from top to bottom). (b) The $m$-pair ($m=0,1,2\text{ and }3$) photon pair creation probabilitystarting from the vacuum state, is shown in the top panels, corresponding to $N=1,2,3,4\text{ and }5$ slabs as in panel (a). In both columns, the lowermost panels show the Floquet frequency versus the wavevector (first Brillouin zone), with the real part of $\omega_F$ in blue and imaginary part in orange.
    }
    \label{fig:twocolumn_fig}
\end{figure}
We now study in detail the dependence of photon transition probabilities on the number of photon pairs created. 
Fig. \ref{fig:twocolumn_fig} shows, for a fixed value of $\alpha=0.5$ and for increasing number of PTC periods ($N = 1, 2, 3, 4, \text{ and }5$), the reflectivity (a) and the transition probabilities for $m=0,1,2 \text{ and } 3$  pairs(b). 
In the lower-most panel of each column we plot the dispersion relation of the PTC, from the Floquet frequencies obtained from the $\boldsymbol{T}-$matrix. We plot $\Re(\omega_F)$ in blue and $ \Im(\omega_F)$ in orange, so that the momentum band gaps can be clearly identified. By looking at the reflectivity plots as $N$ increases (a) and considering the PTC dispersion relation, it is clear that 
the reflectivity exponentially increases with the number of slabs within the momentum bandgaps, for values of $N$ as low as 2 or 3. The non-vanishing imaginary part of the Floquet frequency is responsible for this, and results in larger amplification for $k$ values around $\sim0.75\Omega/c$ due to the larger value of $ \Im(\omega_F)$ in this gap than in the other two shown in the figure. Conversely, within the bands the reflectivity stays lower and the addition of transparency lines where $\mathcal{R}\rightarrow0$ as $N$ increases can be clearly seen. 
\par
Figure \ref{fig:twocolumn_fig}(b) shows photon transition probabilities from the vacuum state calculated with Eqs. \ref{eq:two_mode_squeezing}-\ref{eq:two_mode_squeezing_coeffs}. For all values of $N$, we see that the photon pair creation probability behaves very differently depending on whether $k$ belongs to a band or a gap (see bottom panel for the dispersion relation). Let us first focus in values of momentum that lie within the band of the PTC: it is clear that, independent of $N$, and for a fixed $k$, photon transition probabilities decrease as $m$ increases. However, the particular shape of the probability curves greatly depends on $N$, with all the lines displaying more oscillations as the number of slabs increases, related to the behaviour of the reflectivity in panel (a). In particular, there are a series of $k$ values where the probability of no transition ($m=0$) is maximum (equal to unity), while all the other transitions ($m\ge1$) are zero. These points originate from the transparency conditions of the PTC, $(\boldsymbol{T}^N)_{12}$=0. In between these points, all the photon pair creation probabilities show maxima, with the single pair case ($m=1$) reaching a value of up to $\sim0.25$. However, as we have discussed, the number of transparency lines increases with $N$, 
therefore, sustaining the periodic modulation for a longer time makes the PTC transparent to more waves. Thus, it is not always optimal to have long lasting modulations 
in order to maximize the probability of photon pair creation for a momentum value within the band,
since any mode with real Floquet frequency will eventually pass through the modulation unperturbed if $N$ is large enough, with the PTC becoming transparent to all band modes in the $N\gg1$ limit.

Now focusing on the bandgaps, we see that all the probabilities get progressively squashed towards zero as $N$ increases and become vanishingly small, with seemingly no photons being created inside the momentum gaps for $N\geq3$. However, such a conclusion would ignore that higher order transitions ($m\geq4$) take place and become more probable as the squeezing strengthens (for complex $\omega_F$, $r\sim N$ for $N\gg1$), with lower order ones necessarily becoming less probable. Furthermore, for a given squeezing strength $r$, we have that 
$\text{Prob}(0,0;m,m)=(\tanh^m(r)/\cosh^{m+1}(r))^2 = \mathcal{R}^m/(1+\mathcal{R})^{m+1}$, which decays monotonously with $m$ for fixed $\mathcal{R}$. Also, $\text{Prob}(0,0;m,m)\longrightarrow0$ as $\mathcal{R}\longrightarrow\infty$, such as within the band gaps, for any value of $m$ (a more in-depth discussion is found in \cite{suppmat}). Therefore, the $m$-photon pair creation probability becomes more and more uniform within the bandgaps as more periods are added to the PTC and the generation of pairs of photons becomes asymptotically uniform with $m$ in the $N\gg1$ limit. Hence, the photon number becomes on average very large, 
with all transitions between number states being equally probable, and although
photons are indeed created, there is uncertainty on how many. The quantum statistics of the modes are also discussed in \cite{suppmat}, following \cite{glauber1963quantum} and \cite{kim1989properties}. 
\par

\par
\textit{Conclusions}. We have introduced a theory of photon pair creation in photonic time crystals, and unveiled the connection  between their classical electrodynamical properties -through their reflectivity-, and quantum vacuum amplification processes by means of the squeezing parameter.  Temporal interfaces result in dynamical Casimir processes whereby pairs of forward and backward propagating photons are created from the quantum vacuum in a squeezed state, owing to momentum conservation. Critically, our approach provides a quantum treatment of the PTC regime within an Hermitian framework, allowing to fully treat modes within the momentum bandgaps. We demonstrate that, within these bandgaps, the dynamical Casimir effect is greatly enhanced, and the production of photon pairs further increases and becomes highly fluctuating as more temporal slabs are added to the system. Understanding quantum vacuum amplification in time varying media is important both from a fundamental perspective \cite{nation2012colloquium} and for its practical implications in the generation of quantum light sources \cite{lyubarov2022amplified,dikopoltsev2022light}. 
\begin{acknowledgments}
We acknowledge fruitful conversations with J.B. Pendry, C. Sánchez-Muñoz, E. del Valle and F. Laussy.
We acknowledge funding from: the Spanish Ministry for Science and Innovation (Grant No. RYC2021-031568-I and project No. PID2022-141036NA-I00 financed by MCIN/AEI/10.13039/501100011033 and FSE+) and the CAM Consejería de Educación, Ciencia y Universidades, Viceconsejería de Universidades, Investigación y Ciencia, Dirección General de Investigación e Innovación Tecnológica (Sinergystic Project CAM 2020 NanoQuCo-CM Ref. Y2020/TCS-6545 and CAM FPI grant Ref. A281). PAH further acknowledges funding from Fundaçao para la Cîencia e a Tecnologia under project 2022.06797.PTDC.
\end{acknowledgments}

\bibliography{main}

\begin{thebibliography}{62}%
\makeatletter
\providecommand \@ifxundefined [1]{%
 \@ifx{#1\undefined}
}%
\providecommand \@ifnum [1]{%
 \ifnum #1\expandafter \@firstoftwo
 \else \expandafter \@secondoftwo
 \fi
}%
\providecommand \@ifx [1]{%
 \ifx #1\expandafter \@firstoftwo
 \else \expandafter \@secondoftwo
 \fi
}%
\providecommand \natexlab [1]{#1}%
\providecommand \enquote  [1]{``#1''}%
\providecommand \bibnamefont  [1]{#1}%
\providecommand \bibfnamefont [1]{#1}%
\providecommand \citenamefont [1]{#1}%
\providecommand \href@noop [0]{\@secondoftwo}%
\providecommand \href [0]{\begingroup \@sanitize@url \@href}%
\providecommand \@href[1]{\@@startlink{#1}\@@href}%
\providecommand \@@href[1]{\endgroup#1\@@endlink}%
\providecommand \@sanitize@url [0]{\catcode `\\12\catcode `\$12\catcode `\&12\catcode `\#12\catcode `\^12\catcode `\_12\catcode `\%12\relax}%
\providecommand \@@startlink[1]{}%
\providecommand \@@endlink[0]{}%
\providecommand \url  [0]{\begingroup\@sanitize@url \@url }%
\providecommand \@url [1]{\endgroup\@href {#1}{\urlprefix }}%
\providecommand \urlprefix  [0]{URL }%
\providecommand \Eprint [0]{\href }%
\providecommand \doibase [0]{https://doi.org/}%
\providecommand \selectlanguage [0]{\@gobble}%
\providecommand \bibinfo  [0]{\@secondoftwo}%
\providecommand \bibfield  [0]{\@secondoftwo}%
\providecommand \translation [1]{[#1]}%
\providecommand \BibitemOpen [0]{}%
\providecommand \bibitemStop [0]{}%
\providecommand \bibitemNoStop [0]{.\EOS\space}%
\providecommand \EOS [0]{\spacefactor3000\relax}%
\providecommand \BibitemShut  [1]{\csname bibitem#1\endcsname}%
\let\auto@bib@innerbib\@empty
\bibitem [{\citenamefont {Galiffi}\ \emph {et~al.}(2022)\citenamefont {Galiffi}, \citenamefont {Tirole}, \citenamefont {Yin}, \citenamefont {Li}, \citenamefont {Vezzoli}, \citenamefont {Huidobro}, \citenamefont {Silveirinha}, \citenamefont {Sapienza}, \citenamefont {Al{\`u}},\ and\ \citenamefont {Pendry}}]{galiffi2022photonics}%
  \BibitemOpen
  \bibfield  {author} {\bibinfo {author} {\bibfnamefont {E.}~\bibnamefont {Galiffi}}, \bibinfo {author} {\bibfnamefont {R.}~\bibnamefont {Tirole}}, \bibinfo {author} {\bibfnamefont {S.}~\bibnamefont {Yin}}, \bibinfo {author} {\bibfnamefont {H.}~\bibnamefont {Li}}, \bibinfo {author} {\bibfnamefont {S.}~\bibnamefont {Vezzoli}}, \bibinfo {author} {\bibfnamefont {P.~A.}\ \bibnamefont {Huidobro}}, \bibinfo {author} {\bibfnamefont {M.~G.}\ \bibnamefont {Silveirinha}}, \bibinfo {author} {\bibfnamefont {R.}~\bibnamefont {Sapienza}}, \bibinfo {author} {\bibfnamefont {A.}~\bibnamefont {Al{\`u}}},\ and\ \bibinfo {author} {\bibfnamefont {J.~B.}\ \bibnamefont {Pendry}},\ }\bibfield  {title} {\bibinfo {title} {Photonics of time-varying media},\ }\href@noop {} {\bibfield  {journal} {\bibinfo  {journal} {Advanced Photonics}\ }\textbf {\bibinfo {volume} {4}},\ \bibinfo {pages} {014002} (\bibinfo {year} {2022})}\BibitemShut {NoStop}%
\bibitem [{\citenamefont {Pacheco-Pe{\~n}a}\ \emph {et~al.}(2022)\citenamefont {Pacheco-Pe{\~n}a}, \citenamefont {Sol{\'\i}s},\ and\ \citenamefont {Engheta}}]{pacheco2022time}%
  \BibitemOpen
  \bibfield  {author} {\bibinfo {author} {\bibfnamefont {V.}~\bibnamefont {Pacheco-Pe{\~n}a}}, \bibinfo {author} {\bibfnamefont {D.~M.}\ \bibnamefont {Sol{\'\i}s}},\ and\ \bibinfo {author} {\bibfnamefont {N.}~\bibnamefont {Engheta}},\ }\bibfield  {title} {\bibinfo {title} {Time-varying electromagnetic media: opinion},\ }\href@noop {} {\bibfield  {journal} {\bibinfo  {journal} {Optical Materials Express}\ }\textbf {\bibinfo {volume} {12}},\ \bibinfo {pages} {3829} (\bibinfo {year} {2022})}\BibitemShut {NoStop}%
\bibitem [{\citenamefont {Engheta}(2023)}]{engheta2023four}%
  \BibitemOpen
  \bibfield  {author} {\bibinfo {author} {\bibfnamefont {N.}~\bibnamefont {Engheta}},\ }\bibfield  {title} {\bibinfo {title} {Four-dimensional optics using time-varying metamaterials},\ }\href@noop {} {\bibfield  {journal} {\bibinfo  {journal} {Science}\ }\textbf {\bibinfo {volume} {379}},\ \bibinfo {pages} {1190} (\bibinfo {year} {2023})}\BibitemShut {NoStop}%
\bibitem [{\citenamefont {Boltasseva}\ \emph {et~al.}(2024)\citenamefont {Boltasseva}, \citenamefont {Shalaev},\ and\ \citenamefont {Segev}}]{boltasseva2024photonic}%
  \BibitemOpen
  \bibfield  {author} {\bibinfo {author} {\bibfnamefont {A.}~\bibnamefont {Boltasseva}}, \bibinfo {author} {\bibfnamefont {V.}~\bibnamefont {Shalaev}},\ and\ \bibinfo {author} {\bibfnamefont {M.}~\bibnamefont {Segev}},\ }\bibfield  {title} {\bibinfo {title} {Photonic time crystals: from fundamental insights to novel applications: opinion},\ }\href@noop {} {\bibfield  {journal} {\bibinfo  {journal} {Optical Materials Express}\ }\textbf {\bibinfo {volume} {14}},\ \bibinfo {pages} {592} (\bibinfo {year} {2024})}\BibitemShut {NoStop}%
\bibitem [{\citenamefont {Moussa}\ \emph {et~al.}(2023)\citenamefont {Moussa}, \citenamefont {Xu}, \citenamefont {Yin}, \citenamefont {Galiffi}, \citenamefont {Ra’di},\ and\ \citenamefont {Al{\`u}}}]{moussa2023observation}%
  \BibitemOpen
  \bibfield  {author} {\bibinfo {author} {\bibfnamefont {H.}~\bibnamefont {Moussa}}, \bibinfo {author} {\bibfnamefont {G.}~\bibnamefont {Xu}}, \bibinfo {author} {\bibfnamefont {S.}~\bibnamefont {Yin}}, \bibinfo {author} {\bibfnamefont {E.}~\bibnamefont {Galiffi}}, \bibinfo {author} {\bibfnamefont {Y.}~\bibnamefont {Ra’di}},\ and\ \bibinfo {author} {\bibfnamefont {A.}~\bibnamefont {Al{\`u}}},\ }\bibfield  {title} {\bibinfo {title} {Observation of temporal reflection and broadband frequency translation at photonic time interfaces},\ }\href@noop {} {\bibfield  {journal} {\bibinfo  {journal} {Nature Physics}\ }\textbf {\bibinfo {volume} {19}},\ \bibinfo {pages} {863} (\bibinfo {year} {2023})}\BibitemShut {NoStop}%
\bibitem [{\citenamefont {Zhou}\ \emph {et~al.}(2020)\citenamefont {Zhou}, \citenamefont {Alam}, \citenamefont {Karimi}, \citenamefont {Upham}, \citenamefont {Reshef}, \citenamefont {Liu}, \citenamefont {Willner},\ and\ \citenamefont {Boyd}}]{zhou2020broadband}%
  \BibitemOpen
  \bibfield  {author} {\bibinfo {author} {\bibfnamefont {Y.}~\bibnamefont {Zhou}}, \bibinfo {author} {\bibfnamefont {M.~Z.}\ \bibnamefont {Alam}}, \bibinfo {author} {\bibfnamefont {M.}~\bibnamefont {Karimi}}, \bibinfo {author} {\bibfnamefont {J.}~\bibnamefont {Upham}}, \bibinfo {author} {\bibfnamefont {O.}~\bibnamefont {Reshef}}, \bibinfo {author} {\bibfnamefont {C.}~\bibnamefont {Liu}}, \bibinfo {author} {\bibfnamefont {A.~E.}\ \bibnamefont {Willner}},\ and\ \bibinfo {author} {\bibfnamefont {R.~W.}\ \bibnamefont {Boyd}},\ }\bibfield  {title} {\bibinfo {title} {Broadband frequency translation through time refraction in an epsilon-near-zero material},\ }\href@noop {} {\bibfield  {journal} {\bibinfo  {journal} {Nature communications}\ }\textbf {\bibinfo {volume} {11}},\ \bibinfo {pages} {2180} (\bibinfo {year} {2020})}\BibitemShut {NoStop}%
\bibitem [{\citenamefont {Sol{\'\i}s}\ \emph {et~al.}(2021)\citenamefont {Sol{\'\i}s}, \citenamefont {Kastner},\ and\ \citenamefont {Engheta}}]{solis2021time}%
  \BibitemOpen
  \bibfield  {author} {\bibinfo {author} {\bibfnamefont {D.~M.}\ \bibnamefont {Sol{\'\i}s}}, \bibinfo {author} {\bibfnamefont {R.}~\bibnamefont {Kastner}},\ and\ \bibinfo {author} {\bibfnamefont {N.}~\bibnamefont {Engheta}},\ }\bibfield  {title} {\bibinfo {title} {Time-varying materials in the presence of dispersion: plane-wave propagation in a lorentzian medium with temporal discontinuity},\ }\href@noop {} {\bibfield  {journal} {\bibinfo  {journal} {Photonics Research}\ }\textbf {\bibinfo {volume} {9}},\ \bibinfo {pages} {1842} (\bibinfo {year} {2021})}\BibitemShut {NoStop}%
\bibitem [{\citenamefont {Pang}\ \emph {et~al.}(2021)\citenamefont {Pang}, \citenamefont {Alam}, \citenamefont {Zhou}, \citenamefont {Liu}, \citenamefont {Reshef}, \citenamefont {Manukyan}, \citenamefont {Voegtle}, \citenamefont {Pennathur}, \citenamefont {Tseng}, \citenamefont {Su} \emph {et~al.}}]{pang2021adiabatic}%
  \BibitemOpen
  \bibfield  {author} {\bibinfo {author} {\bibfnamefont {K.}~\bibnamefont {Pang}}, \bibinfo {author} {\bibfnamefont {M.~Z.}\ \bibnamefont {Alam}}, \bibinfo {author} {\bibfnamefont {Y.}~\bibnamefont {Zhou}}, \bibinfo {author} {\bibfnamefont {C.}~\bibnamefont {Liu}}, \bibinfo {author} {\bibfnamefont {O.}~\bibnamefont {Reshef}}, \bibinfo {author} {\bibfnamefont {K.}~\bibnamefont {Manukyan}}, \bibinfo {author} {\bibfnamefont {M.}~\bibnamefont {Voegtle}}, \bibinfo {author} {\bibfnamefont {A.}~\bibnamefont {Pennathur}}, \bibinfo {author} {\bibfnamefont {C.}~\bibnamefont {Tseng}}, \bibinfo {author} {\bibfnamefont {X.}~\bibnamefont {Su}}, \emph {et~al.},\ }\bibfield  {title} {\bibinfo {title} {Adiabatic frequency conversion using a time-varying epsilon-near-zero metasurface},\ }\href@noop {} {\bibfield  {journal} {\bibinfo  {journal} {Nano Letters}\ }\textbf {\bibinfo {volume} {21}},\ \bibinfo {pages} {5907} (\bibinfo {year} {2021})}\BibitemShut {NoStop}%
\bibitem [{\citenamefont {Liu}\ \emph {et~al.}(2021)\citenamefont {Liu}, \citenamefont {Alam}, \citenamefont {Pang}, \citenamefont {Manukyan}, \citenamefont {Reshef}, \citenamefont {Zhou}, \citenamefont {Choudhary}, \citenamefont {Patrow}, \citenamefont {Pennathurs}, \citenamefont {Song} \emph {et~al.}}]{liu2021photon}%
  \BibitemOpen
  \bibfield  {author} {\bibinfo {author} {\bibfnamefont {C.}~\bibnamefont {Liu}}, \bibinfo {author} {\bibfnamefont {M.~Z.}\ \bibnamefont {Alam}}, \bibinfo {author} {\bibfnamefont {K.}~\bibnamefont {Pang}}, \bibinfo {author} {\bibfnamefont {K.}~\bibnamefont {Manukyan}}, \bibinfo {author} {\bibfnamefont {O.}~\bibnamefont {Reshef}}, \bibinfo {author} {\bibfnamefont {Y.}~\bibnamefont {Zhou}}, \bibinfo {author} {\bibfnamefont {S.}~\bibnamefont {Choudhary}}, \bibinfo {author} {\bibfnamefont {J.}~\bibnamefont {Patrow}}, \bibinfo {author} {\bibfnamefont {A.}~\bibnamefont {Pennathurs}}, \bibinfo {author} {\bibfnamefont {H.}~\bibnamefont {Song}}, \emph {et~al.},\ }\bibfield  {title} {\bibinfo {title} {Photon acceleration using a time-varying epsilon-near-zero metasurface},\ }\href@noop {} {\bibfield  {journal} {\bibinfo  {journal} {ACS Photonics}\ }\textbf {\bibinfo {volume} {8}},\ \bibinfo {pages} {716} (\bibinfo {year} {2021})}\BibitemShut {NoStop}%
\bibitem [{\citenamefont {Tirole}\ \emph {et~al.}(2023)\citenamefont {Tirole}, \citenamefont {Vezzoli}, \citenamefont {Galiffi}, \citenamefont {Robertson}, \citenamefont {Maurice}, \citenamefont {Tilmann}, \citenamefont {Maier}, \citenamefont {Pendry},\ and\ \citenamefont {Sapienza}}]{tirole2023double}%
  \BibitemOpen
  \bibfield  {author} {\bibinfo {author} {\bibfnamefont {R.}~\bibnamefont {Tirole}}, \bibinfo {author} {\bibfnamefont {S.}~\bibnamefont {Vezzoli}}, \bibinfo {author} {\bibfnamefont {E.}~\bibnamefont {Galiffi}}, \bibinfo {author} {\bibfnamefont {I.}~\bibnamefont {Robertson}}, \bibinfo {author} {\bibfnamefont {D.}~\bibnamefont {Maurice}}, \bibinfo {author} {\bibfnamefont {B.}~\bibnamefont {Tilmann}}, \bibinfo {author} {\bibfnamefont {S.~A.}\ \bibnamefont {Maier}}, \bibinfo {author} {\bibfnamefont {J.~B.}\ \bibnamefont {Pendry}},\ and\ \bibinfo {author} {\bibfnamefont {R.}~\bibnamefont {Sapienza}},\ }\bibfield  {title} {\bibinfo {title} {Double-slit time diffraction at optical frequencies},\ }\href@noop {} {\bibfield  {journal} {\bibinfo  {journal} {Nature Physics}\ }\textbf {\bibinfo {volume} {19}},\ \bibinfo {pages} {999} (\bibinfo {year} {2023})}\BibitemShut {NoStop}%
\bibitem [{\citenamefont {Huidobro}\ \emph {et~al.}(2019)\citenamefont {Huidobro}, \citenamefont {Galiffi}, \citenamefont {Guenneau}, \citenamefont {Craster},\ and\ \citenamefont {Pendry}}]{huidobro2019fresnel}%
  \BibitemOpen
  \bibfield  {author} {\bibinfo {author} {\bibfnamefont {P.~A.}\ \bibnamefont {Huidobro}}, \bibinfo {author} {\bibfnamefont {E.}~\bibnamefont {Galiffi}}, \bibinfo {author} {\bibfnamefont {S.}~\bibnamefont {Guenneau}}, \bibinfo {author} {\bibfnamefont {R.~V.}\ \bibnamefont {Craster}},\ and\ \bibinfo {author} {\bibfnamefont {J.~B.}\ \bibnamefont {Pendry}},\ }\bibfield  {title} {\bibinfo {title} {Fresnel drag in space--time-modulated metamaterials},\ }\href@noop {} {\bibfield  {journal} {\bibinfo  {journal} {Proceedings of the National Academy of Sciences}\ }\textbf {\bibinfo {volume} {116}},\ \bibinfo {pages} {24943} (\bibinfo {year} {2019})}\BibitemShut {NoStop}%
\bibitem [{\citenamefont {Prud{\^e}ncio}\ and\ \citenamefont {Silveirinha}(2023)}]{prudencio2023replicating}%
  \BibitemOpen
  \bibfield  {author} {\bibinfo {author} {\bibfnamefont {F.~R.}\ \bibnamefont {Prud{\^e}ncio}}\ and\ \bibinfo {author} {\bibfnamefont {M.~G.}\ \bibnamefont {Silveirinha}},\ }\bibfield  {title} {\bibinfo {title} {Replicating physical motion with minkowskian isorefractive spacetime crystals},\ }\href@noop {} {\bibfield  {journal} {\bibinfo  {journal} {Nanophotonics}\ }\textbf {\bibinfo {volume} {12}},\ \bibinfo {pages} {3007} (\bibinfo {year} {2023})}\BibitemShut {NoStop}%
\bibitem [{\citenamefont {Wang}\ \emph {et~al.}(2020)\citenamefont {Wang}, \citenamefont {Diaz-Rubio}, \citenamefont {Li}, \citenamefont {Tretyakov},\ and\ \citenamefont {Alu}}]{wang2020theory}%
  \BibitemOpen
  \bibfield  {author} {\bibinfo {author} {\bibfnamefont {X.}~\bibnamefont {Wang}}, \bibinfo {author} {\bibfnamefont {A.}~\bibnamefont {Diaz-Rubio}}, \bibinfo {author} {\bibfnamefont {H.}~\bibnamefont {Li}}, \bibinfo {author} {\bibfnamefont {S.~A.}\ \bibnamefont {Tretyakov}},\ and\ \bibinfo {author} {\bibfnamefont {A.}~\bibnamefont {Alu}},\ }\bibfield  {title} {\bibinfo {title} {Theory and design of multifunctional space-time metasurfaces},\ }\href@noop {} {\bibfield  {journal} {\bibinfo  {journal} {Physical Review Applied}\ }\textbf {\bibinfo {volume} {13}},\ \bibinfo {pages} {044040} (\bibinfo {year} {2020})}\BibitemShut {NoStop}%
\bibitem [{\citenamefont {Li}\ \emph {et~al.}(2023)\citenamefont {Li}, \citenamefont {Yin}, \citenamefont {He}, \citenamefont {Xu}, \citenamefont {Al{\`u}},\ and\ \citenamefont {Shapiro}}]{li2023stationary}%
  \BibitemOpen
  \bibfield  {author} {\bibinfo {author} {\bibfnamefont {H.}~\bibnamefont {Li}}, \bibinfo {author} {\bibfnamefont {S.}~\bibnamefont {Yin}}, \bibinfo {author} {\bibfnamefont {H.}~\bibnamefont {He}}, \bibinfo {author} {\bibfnamefont {J.}~\bibnamefont {Xu}}, \bibinfo {author} {\bibfnamefont {A.}~\bibnamefont {Al{\`u}}},\ and\ \bibinfo {author} {\bibfnamefont {B.}~\bibnamefont {Shapiro}},\ }\bibfield  {title} {\bibinfo {title} {Stationary charge radiation in anisotropic photonic time crystals},\ }\href@noop {} {\bibfield  {journal} {\bibinfo  {journal} {Physical Review Letters}\ }\textbf {\bibinfo {volume} {130}},\ \bibinfo {pages} {093803} (\bibinfo {year} {2023})}\BibitemShut {NoStop}%
\bibitem [{\citenamefont {Chamanara}\ \emph {et~al.}(2019)\citenamefont {Chamanara}, \citenamefont {Vahabzadeh},\ and\ \citenamefont {Caloz}}]{chamanara2019simultaneous}%
  \BibitemOpen
  \bibfield  {author} {\bibinfo {author} {\bibfnamefont {N.}~\bibnamefont {Chamanara}}, \bibinfo {author} {\bibfnamefont {Y.}~\bibnamefont {Vahabzadeh}},\ and\ \bibinfo {author} {\bibfnamefont {C.}~\bibnamefont {Caloz}},\ }\bibfield  {title} {\bibinfo {title} {Simultaneous control of the spatial and temporal spectra of light with space-time varying metasurfaces},\ }\href@noop {} {\bibfield  {journal} {\bibinfo  {journal} {IEEE Transactions on Antennas and Propagation}\ }\textbf {\bibinfo {volume} {67}},\ \bibinfo {pages} {2430} (\bibinfo {year} {2019})}\BibitemShut {NoStop}%
\bibitem [{\citenamefont {Galiffi}\ \emph {et~al.}(2023)\citenamefont {Galiffi}, \citenamefont {Xu}, \citenamefont {Yin}, \citenamefont {Moussa}, \citenamefont {Ra’di},\ and\ \citenamefont {Al{\`u}}}]{galiffi2023broadband}%
  \BibitemOpen
  \bibfield  {author} {\bibinfo {author} {\bibfnamefont {E.}~\bibnamefont {Galiffi}}, \bibinfo {author} {\bibfnamefont {G.}~\bibnamefont {Xu}}, \bibinfo {author} {\bibfnamefont {S.}~\bibnamefont {Yin}}, \bibinfo {author} {\bibfnamefont {H.}~\bibnamefont {Moussa}}, \bibinfo {author} {\bibfnamefont {Y.}~\bibnamefont {Ra’di}},\ and\ \bibinfo {author} {\bibfnamefont {A.}~\bibnamefont {Al{\`u}}},\ }\bibfield  {title} {\bibinfo {title} {Broadband coherent wave control through photonic collisions at time interfaces},\ }\href@noop {} {\bibfield  {journal} {\bibinfo  {journal} {Nature Physics}\ }\textbf {\bibinfo {volume} {19}},\ \bibinfo {pages} {1703} (\bibinfo {year} {2023})}\BibitemShut {NoStop}%
\bibitem [{\citenamefont {Prud\^encio}\ and\ \citenamefont {Silveirinha}(2023)}]{prudencio2023synthetic}%
  \BibitemOpen
  \bibfield  {author} {\bibinfo {author} {\bibfnamefont {F.~R.}\ \bibnamefont {Prud\^encio}}\ and\ \bibinfo {author} {\bibfnamefont {M.~G.}\ \bibnamefont {Silveirinha}},\ }\bibfield  {title} {\bibinfo {title} {Synthetic axion response with space-time crystals},\ }\href {https://doi.org/10.1103/PhysRevApplied.19.024031} {\bibfield  {journal} {\bibinfo  {journal} {Phys. Rev. Appl.}\ }\textbf {\bibinfo {volume} {19}},\ \bibinfo {pages} {024031} (\bibinfo {year} {2023})}\BibitemShut {NoStop}%
\bibitem [{\citenamefont {Pendry}\ \emph {et~al.}(2022)\citenamefont {Pendry}, \citenamefont {Galiffi},\ and\ \citenamefont {Huidobro}}]{pendry2022photon}%
  \BibitemOpen
  \bibfield  {author} {\bibinfo {author} {\bibfnamefont {J.}~\bibnamefont {Pendry}}, \bibinfo {author} {\bibfnamefont {E.}~\bibnamefont {Galiffi}},\ and\ \bibinfo {author} {\bibfnamefont {P.}~\bibnamefont {Huidobro}},\ }\bibfield  {title} {\bibinfo {title} {Photon conservation in trans-luminal metamaterials},\ }\href@noop {} {\bibfield  {journal} {\bibinfo  {journal} {Optica}\ }\textbf {\bibinfo {volume} {9}},\ \bibinfo {pages} {724} (\bibinfo {year} {2022})}\BibitemShut {NoStop}%
\bibitem [{\citenamefont {Bahrami}\ \emph {et~al.}(2023)\citenamefont {Bahrami}, \citenamefont {Deck-L{\'e}ger},\ and\ \citenamefont {Caloz}}]{bahrami2023electrodynamics}%
  \BibitemOpen
  \bibfield  {author} {\bibinfo {author} {\bibfnamefont {A.}~\bibnamefont {Bahrami}}, \bibinfo {author} {\bibfnamefont {Z.-L.}\ \bibnamefont {Deck-L{\'e}ger}},\ and\ \bibinfo {author} {\bibfnamefont {C.}~\bibnamefont {Caloz}},\ }\bibfield  {title} {\bibinfo {title} {Electrodynamics of accelerated-modulation space-time metamaterials},\ }\href@noop {} {\bibfield  {journal} {\bibinfo  {journal} {Physical Review Applied}\ }\textbf {\bibinfo {volume} {19}},\ \bibinfo {pages} {054044} (\bibinfo {year} {2023})}\BibitemShut {NoStop}%
\bibitem [{\citenamefont {Horsley}\ and\ \citenamefont {Pendry}(2023)}]{horsley2023quantum}%
  \BibitemOpen
  \bibfield  {author} {\bibinfo {author} {\bibfnamefont {S.~A.}\ \bibnamefont {Horsley}}\ and\ \bibinfo {author} {\bibfnamefont {J.~B.}\ \bibnamefont {Pendry}},\ }\bibfield  {title} {\bibinfo {title} {Quantum electrodynamics of time-varying gratings},\ }\href@noop {} {\bibfield  {journal} {\bibinfo  {journal} {Proceedings of the National Academy of Sciences}\ }\textbf {\bibinfo {volume} {120}},\ \bibinfo {pages} {e2302652120} (\bibinfo {year} {2023})}\BibitemShut {NoStop}%
\bibitem [{\citenamefont {Morgenthaler}(1958)}]{morgenthaler1958velocity}%
  \BibitemOpen
  \bibfield  {author} {\bibinfo {author} {\bibfnamefont {F.~R.}\ \bibnamefont {Morgenthaler}},\ }\bibfield  {title} {\bibinfo {title} {Velocity modulation of electromagnetic waves},\ }\href@noop {} {\bibfield  {journal} {\bibinfo  {journal} {IRE Transactions on microwave theory and techniques}\ }\textbf {\bibinfo {volume} {6}},\ \bibinfo {pages} {167} (\bibinfo {year} {1958})}\BibitemShut {NoStop}%
\bibitem [{\citenamefont {Felsen}\ and\ \citenamefont {Whitman}(1970)}]{felsen1970wave}%
  \BibitemOpen
  \bibfield  {author} {\bibinfo {author} {\bibfnamefont {L.}~\bibnamefont {Felsen}}\ and\ \bibinfo {author} {\bibfnamefont {G.}~\bibnamefont {Whitman}},\ }\bibfield  {title} {\bibinfo {title} {Wave propagation in time-varying media},\ }\href@noop {} {\bibfield  {journal} {\bibinfo  {journal} {IEEE Transactions on Antennas and Propagation}\ }\textbf {\bibinfo {volume} {18}},\ \bibinfo {pages} {242} (\bibinfo {year} {1970})}\BibitemShut {NoStop}%
\bibitem [{\citenamefont {Fante}(1971)}]{fante1971transmission}%
  \BibitemOpen
  \bibfield  {author} {\bibinfo {author} {\bibfnamefont {R.}~\bibnamefont {Fante}},\ }\bibfield  {title} {\bibinfo {title} {Transmission of electromagnetic waves into time-varying media},\ }\href@noop {} {\bibfield  {journal} {\bibinfo  {journal} {IEEE Transactions on Antennas and Propagation}\ }\textbf {\bibinfo {volume} {19}},\ \bibinfo {pages} {417} (\bibinfo {year} {1971})}\BibitemShut {NoStop}%
\bibitem [{\citenamefont {Ramaccia}\ \emph {et~al.}(2020)\citenamefont {Ramaccia}, \citenamefont {Toscano},\ and\ \citenamefont {Bilotti}}]{Ramaccia:20}%
  \BibitemOpen
  \bibfield  {author} {\bibinfo {author} {\bibfnamefont {D.}~\bibnamefont {Ramaccia}}, \bibinfo {author} {\bibfnamefont {A.}~\bibnamefont {Toscano}},\ and\ \bibinfo {author} {\bibfnamefont {F.}~\bibnamefont {Bilotti}},\ }\bibfield  {title} {\bibinfo {title} {Light propagation through metamaterial temporal slabs: reflection, refraction, and special cases},\ }\href@noop {} {\bibfield  {journal} {\bibinfo  {journal} {Opt. Lett.}\ }\textbf {\bibinfo {volume} {45}},\ \bibinfo {pages} {5836} (\bibinfo {year} {2020})}\BibitemShut {NoStop}%
\bibitem [{\citenamefont {Ramaccia}\ \emph {et~al.}(2021)\citenamefont {Ramaccia}, \citenamefont {Al{\`u}}, \citenamefont {Toscano},\ and\ \citenamefont {Bilotti}}]{ramaccia2021temporal}%
  \BibitemOpen
  \bibfield  {author} {\bibinfo {author} {\bibfnamefont {D.}~\bibnamefont {Ramaccia}}, \bibinfo {author} {\bibfnamefont {A.}~\bibnamefont {Al{\`u}}}, \bibinfo {author} {\bibfnamefont {A.}~\bibnamefont {Toscano}},\ and\ \bibinfo {author} {\bibfnamefont {F.}~\bibnamefont {Bilotti}},\ }\bibfield  {title} {\bibinfo {title} {Temporal multilayer structures for designing higher-order transfer functions using time-varying metamaterials},\ }\href@noop {} {\bibfield  {journal} {\bibinfo  {journal} {Applied Physics Letters}\ }\textbf {\bibinfo {volume} {118}} (\bibinfo {year} {2021})}\BibitemShut {NoStop}%
\bibitem [{\citenamefont {Mirmoosa}\ \emph {et~al.}(2024)\citenamefont {Mirmoosa}, \citenamefont {Mostafa}, \citenamefont {Norrman},\ and\ \citenamefont {Tretyakov}}]{mirmoosa2024time}%
  \BibitemOpen
  \bibfield  {author} {\bibinfo {author} {\bibfnamefont {M.}~\bibnamefont {Mirmoosa}}, \bibinfo {author} {\bibfnamefont {M.}~\bibnamefont {Mostafa}}, \bibinfo {author} {\bibfnamefont {A.}~\bibnamefont {Norrman}},\ and\ \bibinfo {author} {\bibfnamefont {S.}~\bibnamefont {Tretyakov}},\ }\bibfield  {title} {\bibinfo {title} {Time interfaces in bianisotropic media},\ }\href@noop {} {\bibfield  {journal} {\bibinfo  {journal} {Physical Review Research}\ }\textbf {\bibinfo {volume} {6}},\ \bibinfo {pages} {013334} (\bibinfo {year} {2024})}\BibitemShut {NoStop}%
\bibitem [{\citenamefont {Mart{\'\i}nez-Romero}\ and\ \citenamefont {Halevi}(2018)}]{martinez2018parametric}%
  \BibitemOpen
  \bibfield  {author} {\bibinfo {author} {\bibfnamefont {J.~S.}\ \bibnamefont {Mart{\'\i}nez-Romero}}\ and\ \bibinfo {author} {\bibfnamefont {P.}~\bibnamefont {Halevi}},\ }\bibfield  {title} {\bibinfo {title} {Parametric resonances in a temporal photonic crystal slab},\ }\href@noop {} {\bibfield  {journal} {\bibinfo  {journal} {Physical Review A}\ }\textbf {\bibinfo {volume} {98}},\ \bibinfo {pages} {053852} (\bibinfo {year} {2018})}\BibitemShut {NoStop}%
\bibitem [{\citenamefont {Apffel}\ and\ \citenamefont {Fort}(2022)}]{apffel2022frequency}%
  \BibitemOpen
  \bibfield  {author} {\bibinfo {author} {\bibfnamefont {B.}~\bibnamefont {Apffel}}\ and\ \bibinfo {author} {\bibfnamefont {E.}~\bibnamefont {Fort}},\ }\bibfield  {title} {\bibinfo {title} {Frequency conversion cascade by crossing multiple space and time interfaces},\ }\href@noop {} {\bibfield  {journal} {\bibinfo  {journal} {Physical Review Letters}\ }\textbf {\bibinfo {volume} {128}},\ \bibinfo {pages} {064501} (\bibinfo {year} {2022})}\BibitemShut {NoStop}%
\bibitem [{\citenamefont {Bacot}\ \emph {et~al.}(2016)\citenamefont {Bacot}, \citenamefont {Labousse}, \citenamefont {Eddi}, \citenamefont {Fink},\ and\ \citenamefont {Fort}}]{bacot2016time}%
  \BibitemOpen
  \bibfield  {author} {\bibinfo {author} {\bibfnamefont {V.}~\bibnamefont {Bacot}}, \bibinfo {author} {\bibfnamefont {M.}~\bibnamefont {Labousse}}, \bibinfo {author} {\bibfnamefont {A.}~\bibnamefont {Eddi}}, \bibinfo {author} {\bibfnamefont {M.}~\bibnamefont {Fink}},\ and\ \bibinfo {author} {\bibfnamefont {E.}~\bibnamefont {Fort}},\ }\bibfield  {title} {\bibinfo {title} {Time reversal and holography with spacetime transformations},\ }\href@noop {} {\bibfield  {journal} {\bibinfo  {journal} {Nature Physics}\ }\textbf {\bibinfo {volume} {12}},\ \bibinfo {pages} {972} (\bibinfo {year} {2016})}\BibitemShut {NoStop}%
\bibitem [{\citenamefont {Fink}\ and\ \citenamefont {Fort}(2017)}]{fink2017time}%
  \BibitemOpen
  \bibfield  {author} {\bibinfo {author} {\bibfnamefont {M.}~\bibnamefont {Fink}}\ and\ \bibinfo {author} {\bibfnamefont {E.}~\bibnamefont {Fort}},\ }\bibfield  {title} {\bibinfo {title} {From the time-reversal mirror to the instantaneous time mirror},\ }\href@noop {} {\bibfield  {journal} {\bibinfo  {journal} {The European Physical Journal Special Topics}\ }\textbf {\bibinfo {volume} {226}},\ \bibinfo {pages} {1477} (\bibinfo {year} {2017})}\BibitemShut {NoStop}%
\bibitem [{\citenamefont {Pacheco-Pe{\~n}a}\ and\ \citenamefont {Engheta}(2020)}]{pacheco2020antireflection}%
  \BibitemOpen
  \bibfield  {author} {\bibinfo {author} {\bibfnamefont {V.}~\bibnamefont {Pacheco-Pe{\~n}a}}\ and\ \bibinfo {author} {\bibfnamefont {N.}~\bibnamefont {Engheta}},\ }\bibfield  {title} {\bibinfo {title} {Antireflection temporal coatings},\ }\href@noop {} {\bibfield  {journal} {\bibinfo  {journal} {Optica}\ }\textbf {\bibinfo {volume} {7}},\ \bibinfo {pages} {323} (\bibinfo {year} {2020})}\BibitemShut {NoStop}%
\bibitem [{\citenamefont {Reyes-Ayona}\ and\ \citenamefont {Halevi}(2015)}]{reyes2015observation}%
  \BibitemOpen
  \bibfield  {author} {\bibinfo {author} {\bibfnamefont {J.}~\bibnamefont {Reyes-Ayona}}\ and\ \bibinfo {author} {\bibfnamefont {P.}~\bibnamefont {Halevi}},\ }\bibfield  {title} {\bibinfo {title} {Observation of genuine wave vector (k or $\beta$) gap in a dynamic transmission line and temporal photonic crystals},\ }\href@noop {} {\bibfield  {journal} {\bibinfo  {journal} {Applied Physics Letters}\ }\textbf {\bibinfo {volume} {107}} (\bibinfo {year} {2015})}\BibitemShut {NoStop}%
\bibitem [{\citenamefont {Wang}\ \emph {et~al.}(2023)\citenamefont {Wang}, \citenamefont {Mirmoosa}, \citenamefont {Asadchy}, \citenamefont {Rockstuhl}, \citenamefont {Fan},\ and\ \citenamefont {Tretyakov}}]{wang2023metasurface}%
  \BibitemOpen
  \bibfield  {author} {\bibinfo {author} {\bibfnamefont {X.}~\bibnamefont {Wang}}, \bibinfo {author} {\bibfnamefont {M.~S.}\ \bibnamefont {Mirmoosa}}, \bibinfo {author} {\bibfnamefont {V.~S.}\ \bibnamefont {Asadchy}}, \bibinfo {author} {\bibfnamefont {C.}~\bibnamefont {Rockstuhl}}, \bibinfo {author} {\bibfnamefont {S.}~\bibnamefont {Fan}},\ and\ \bibinfo {author} {\bibfnamefont {S.~A.}\ \bibnamefont {Tretyakov}},\ }\bibfield  {title} {\bibinfo {title} {Metasurface-based realization of photonic time crystals},\ }\href@noop {} {\bibfield  {journal} {\bibinfo  {journal} {Science Advances}\ }\textbf {\bibinfo {volume} {9}},\ \bibinfo {pages} {eadg7541} (\bibinfo {year} {2023})}\BibitemShut {NoStop}%
\bibitem [{\citenamefont {Liberal}\ and\ \citenamefont {Engheta}(2017)}]{liberal2017near}%
  \BibitemOpen
  \bibfield  {author} {\bibinfo {author} {\bibfnamefont {I.}~\bibnamefont {Liberal}}\ and\ \bibinfo {author} {\bibfnamefont {N.}~\bibnamefont {Engheta}},\ }\bibfield  {title} {\bibinfo {title} {Near-zero refractive index photonics},\ }\href@noop {} {\bibfield  {journal} {\bibinfo  {journal} {Nature Photonics}\ }\textbf {\bibinfo {volume} {11}},\ \bibinfo {pages} {149} (\bibinfo {year} {2017})}\BibitemShut {NoStop}%
\bibitem [{\citenamefont {Boyd}\ \emph {et~al.}(2017)\citenamefont {Boyd}, \citenamefont {Reshef}, \citenamefont {Giese}, \citenamefont {Alam}, \citenamefont {Upham},\ and\ \citenamefont {De~Leon}}]{boyd2017beyond}%
  \BibitemOpen
  \bibfield  {author} {\bibinfo {author} {\bibfnamefont {R.~W.}\ \bibnamefont {Boyd}}, \bibinfo {author} {\bibfnamefont {O.}~\bibnamefont {Reshef}}, \bibinfo {author} {\bibfnamefont {E.}~\bibnamefont {Giese}}, \bibinfo {author} {\bibfnamefont {M.~Z.}\ \bibnamefont {Alam}}, \bibinfo {author} {\bibfnamefont {J.}~\bibnamefont {Upham}},\ and\ \bibinfo {author} {\bibfnamefont {I.}~\bibnamefont {De~Leon}},\ }\bibfield  {title} {\bibinfo {title} {Beyond the perturbative description of the nonlinear optical response of highly nonlinear, epsilon-near-zero materials},\ }in\ \href@noop {} {\emph {\bibinfo {booktitle} {Nonlinear Optics}}}\ (\bibinfo {organization} {Optica Publishing Group},\ \bibinfo {year} {2017})\ pp.\ \bibinfo {pages} {NW1A--2}\BibitemShut {NoStop}%
\bibitem [{\citenamefont {Reshef}\ \emph {et~al.}(2019)\citenamefont {Reshef}, \citenamefont {De~Leon}, \citenamefont {Alam},\ and\ \citenamefont {Boyd}}]{reshef2019nonlinear}%
  \BibitemOpen
  \bibfield  {author} {\bibinfo {author} {\bibfnamefont {O.}~\bibnamefont {Reshef}}, \bibinfo {author} {\bibfnamefont {I.}~\bibnamefont {De~Leon}}, \bibinfo {author} {\bibfnamefont {M.~Z.}\ \bibnamefont {Alam}},\ and\ \bibinfo {author} {\bibfnamefont {R.~W.}\ \bibnamefont {Boyd}},\ }\bibfield  {title} {\bibinfo {title} {Nonlinear optical effects in epsilon-near-zero media},\ }\href@noop {} {\bibfield  {journal} {\bibinfo  {journal} {Nature Reviews Materials}\ }\textbf {\bibinfo {volume} {4}},\ \bibinfo {pages} {535} (\bibinfo {year} {2019})}\BibitemShut {NoStop}%
\bibitem [{\citenamefont {Bohn}\ \emph {et~al.}(2021)\citenamefont {Bohn}, \citenamefont {Luk}, \citenamefont {Tollerton}, \citenamefont {Hutchings}, \citenamefont {Brener}, \citenamefont {Horsley}, \citenamefont {Barnes},\ and\ \citenamefont {Hendry}}]{bohn2021all}%
  \BibitemOpen
  \bibfield  {author} {\bibinfo {author} {\bibfnamefont {J.}~\bibnamefont {Bohn}}, \bibinfo {author} {\bibfnamefont {T.~S.}\ \bibnamefont {Luk}}, \bibinfo {author} {\bibfnamefont {C.}~\bibnamefont {Tollerton}}, \bibinfo {author} {\bibfnamefont {S.~W.}\ \bibnamefont {Hutchings}}, \bibinfo {author} {\bibfnamefont {I.}~\bibnamefont {Brener}}, \bibinfo {author} {\bibfnamefont {S.}~\bibnamefont {Horsley}}, \bibinfo {author} {\bibfnamefont {W.~L.}\ \bibnamefont {Barnes}},\ and\ \bibinfo {author} {\bibfnamefont {E.}~\bibnamefont {Hendry}},\ }\bibfield  {title} {\bibinfo {title} {All-optical switching of an epsilon-near-zero plasmon resonance in indium tin oxide},\ }\href@noop {} {\bibfield  {journal} {\bibinfo  {journal} {Nature communications}\ }\textbf {\bibinfo {volume} {12}},\ \bibinfo {pages} {1017} (\bibinfo {year} {2021})}\BibitemShut {NoStop}%
\bibitem [{\citenamefont {Vezzoli}\ \emph {et~al.}(2018)\citenamefont {Vezzoli}, \citenamefont {Bruno}, \citenamefont {DeVault}, \citenamefont {Roger}, \citenamefont {Shalaev}, \citenamefont {Boltasseva}, \citenamefont {Ferrera}, \citenamefont {Clerici}, \citenamefont {Dubietis},\ and\ \citenamefont {Faccio}}]{vezzoli2018optical}%
  \BibitemOpen
  \bibfield  {author} {\bibinfo {author} {\bibfnamefont {S.}~\bibnamefont {Vezzoli}}, \bibinfo {author} {\bibfnamefont {V.}~\bibnamefont {Bruno}}, \bibinfo {author} {\bibfnamefont {C.}~\bibnamefont {DeVault}}, \bibinfo {author} {\bibfnamefont {T.}~\bibnamefont {Roger}}, \bibinfo {author} {\bibfnamefont {V.~M.}\ \bibnamefont {Shalaev}}, \bibinfo {author} {\bibfnamefont {A.}~\bibnamefont {Boltasseva}}, \bibinfo {author} {\bibfnamefont {M.}~\bibnamefont {Ferrera}}, \bibinfo {author} {\bibfnamefont {M.}~\bibnamefont {Clerici}}, \bibinfo {author} {\bibfnamefont {A.}~\bibnamefont {Dubietis}},\ and\ \bibinfo {author} {\bibfnamefont {D.}~\bibnamefont {Faccio}},\ }\bibfield  {title} {\bibinfo {title} {Optical time reversal from time-dependent epsilon-near-zero media},\ }\href@noop {} {\bibfield  {journal} {\bibinfo  {journal} {Physical review letters}\ }\textbf {\bibinfo {volume} {120}},\ \bibinfo {pages} {043902} (\bibinfo {year} {2018})}\BibitemShut {NoStop}%
\bibitem [{\citenamefont {Tirole}\ \emph {et~al.}(2022)\citenamefont {Tirole}, \citenamefont {Galiffi}, \citenamefont {Dranczewski}, \citenamefont {Attavar}, \citenamefont {Tilmann}, \citenamefont {Wang}, \citenamefont {Huidobro}, \citenamefont {Al{\'u}}, \citenamefont {Pendry}, \citenamefont {Maier} \emph {et~al.}}]{tirole2022saturable}%
  \BibitemOpen
  \bibfield  {author} {\bibinfo {author} {\bibfnamefont {R.}~\bibnamefont {Tirole}}, \bibinfo {author} {\bibfnamefont {E.}~\bibnamefont {Galiffi}}, \bibinfo {author} {\bibfnamefont {J.}~\bibnamefont {Dranczewski}}, \bibinfo {author} {\bibfnamefont {T.}~\bibnamefont {Attavar}}, \bibinfo {author} {\bibfnamefont {B.}~\bibnamefont {Tilmann}}, \bibinfo {author} {\bibfnamefont {Y.-T.}\ \bibnamefont {Wang}}, \bibinfo {author} {\bibfnamefont {P.~A.}\ \bibnamefont {Huidobro}}, \bibinfo {author} {\bibfnamefont {A.}~\bibnamefont {Al{\'u}}}, \bibinfo {author} {\bibfnamefont {J.~B.}\ \bibnamefont {Pendry}}, \bibinfo {author} {\bibfnamefont {S.~A.}\ \bibnamefont {Maier}}, \emph {et~al.},\ }\bibfield  {title} {\bibinfo {title} {Saturable time-varying mirror based on an epsilon-near-zero material},\ }\href@noop {} {\bibfield  {journal} {\bibinfo  {journal} {Physical Review Applied}\ }\textbf {\bibinfo {volume} {18}},\ \bibinfo {pages} {054067} (\bibinfo {year} {2022})}\BibitemShut {NoStop}%
\bibitem [{\citenamefont {Lustig}\ \emph {et~al.}(2023)\citenamefont {Lustig}, \citenamefont {Segal}, \citenamefont {Saha}, \citenamefont {Bordo}, \citenamefont {Chowdhury}, \citenamefont {Sharabi}, \citenamefont {Fleischer}, \citenamefont {Boltasseva}, \citenamefont {Cohen}, \citenamefont {Shalaev} \emph {et~al.}}]{lustig2023time}%
  \BibitemOpen
  \bibfield  {author} {\bibinfo {author} {\bibfnamefont {E.}~\bibnamefont {Lustig}}, \bibinfo {author} {\bibfnamefont {O.}~\bibnamefont {Segal}}, \bibinfo {author} {\bibfnamefont {S.}~\bibnamefont {Saha}}, \bibinfo {author} {\bibfnamefont {E.}~\bibnamefont {Bordo}}, \bibinfo {author} {\bibfnamefont {S.~N.}\ \bibnamefont {Chowdhury}}, \bibinfo {author} {\bibfnamefont {Y.}~\bibnamefont {Sharabi}}, \bibinfo {author} {\bibfnamefont {A.}~\bibnamefont {Fleischer}}, \bibinfo {author} {\bibfnamefont {A.}~\bibnamefont {Boltasseva}}, \bibinfo {author} {\bibfnamefont {O.}~\bibnamefont {Cohen}}, \bibinfo {author} {\bibfnamefont {V.~M.}\ \bibnamefont {Shalaev}}, \emph {et~al.},\ }\bibfield  {title} {\bibinfo {title} {Time-refraction optics with single cycle modulation},\ }\href@noop {} {\bibfield  {journal} {\bibinfo  {journal} {Nanophotonics}\ }\textbf {\bibinfo {volume} {12}},\ \bibinfo {pages} {2221} (\bibinfo {year} {2023})}\BibitemShut {NoStop}%
\bibitem [{\citenamefont {Mendon{\c{c}}a}\ \emph {et~al.}(2000)\citenamefont {Mendon{\c{c}}a}, \citenamefont {Guerreiro},\ and\ \citenamefont {Martins}}]{mendoncca2000quantum}%
  \BibitemOpen
  \bibfield  {author} {\bibinfo {author} {\bibfnamefont {J.}~\bibnamefont {Mendon{\c{c}}a}}, \bibinfo {author} {\bibfnamefont {A.}~\bibnamefont {Guerreiro}},\ and\ \bibinfo {author} {\bibfnamefont {A.~M.}\ \bibnamefont {Martins}},\ }\bibfield  {title} {\bibinfo {title} {Quantum theory of time refraction},\ }\href@noop {} {\bibfield  {journal} {\bibinfo  {journal} {Physical Review A}\ }\textbf {\bibinfo {volume} {62}},\ \bibinfo {pages} {033805} (\bibinfo {year} {2000})}\BibitemShut {NoStop}%
\bibitem [{\citenamefont {Mendon{\c{c}}a}\ \emph {et~al.}(2003)\citenamefont {Mendon{\c{c}}a}, \citenamefont {Martins},\ and\ \citenamefont {Guerreiro}}]{mendoncca2003temporal}%
  \BibitemOpen
  \bibfield  {author} {\bibinfo {author} {\bibfnamefont {J.}~\bibnamefont {Mendon{\c{c}}a}}, \bibinfo {author} {\bibfnamefont {A.}~\bibnamefont {Martins}},\ and\ \bibinfo {author} {\bibfnamefont {A.}~\bibnamefont {Guerreiro}},\ }\bibfield  {title} {\bibinfo {title} {Temporal beam splitter and temporal interference},\ }\href@noop {} {\bibfield  {journal} {\bibinfo  {journal} {Physical Review A}\ }\textbf {\bibinfo {volume} {68}},\ \bibinfo {pages} {043801} (\bibinfo {year} {2003})}\BibitemShut {NoStop}%
\bibitem [{\citenamefont {Lyubarov}\ \emph {et~al.}(2022)\citenamefont {Lyubarov}, \citenamefont {Lumer}, \citenamefont {Dikopoltsev}, \citenamefont {Lustig}, \citenamefont {Sharabi},\ and\ \citenamefont {Segev}}]{lyubarov2022amplified}%
  \BibitemOpen
  \bibfield  {author} {\bibinfo {author} {\bibfnamefont {M.}~\bibnamefont {Lyubarov}}, \bibinfo {author} {\bibfnamefont {Y.}~\bibnamefont {Lumer}}, \bibinfo {author} {\bibfnamefont {A.}~\bibnamefont {Dikopoltsev}}, \bibinfo {author} {\bibfnamefont {E.}~\bibnamefont {Lustig}}, \bibinfo {author} {\bibfnamefont {Y.}~\bibnamefont {Sharabi}},\ and\ \bibinfo {author} {\bibfnamefont {M.}~\bibnamefont {Segev}},\ }\bibfield  {title} {\bibinfo {title} {Amplified emission and lasing in photonic time crystals},\ }\href@noop {} {\bibfield  {journal} {\bibinfo  {journal} {Science}\ }\textbf {\bibinfo {volume} {377}},\ \bibinfo {pages} {425} (\bibinfo {year} {2022})}\BibitemShut {NoStop}%
\bibitem [{\citenamefont {V{\'a}zquez-Lozano}\ and\ \citenamefont {Liberal}(2023)}]{vazquez2023shaping}%
  \BibitemOpen
  \bibfield  {author} {\bibinfo {author} {\bibfnamefont {J.~E.}\ \bibnamefont {V{\'a}zquez-Lozano}}\ and\ \bibinfo {author} {\bibfnamefont {I.}~\bibnamefont {Liberal}},\ }\bibfield  {title} {\bibinfo {title} {Shaping the quantum vacuum with anisotropic temporal boundaries},\ }\href@noop {} {\bibfield  {journal} {\bibinfo  {journal} {Nanophotonics}\ }\textbf {\bibinfo {volume} {12}},\ \bibinfo {pages} {539} (\bibinfo {year} {2023})}\BibitemShut {NoStop}%
\bibitem [{\citenamefont {Liberal}\ \emph {et~al.}(2023)\citenamefont {Liberal}, \citenamefont {V{\'a}zquez-Lozano},\ and\ \citenamefont {Pacheco-Pe{\~n}a}}]{liberal2023quantum}%
  \BibitemOpen
  \bibfield  {author} {\bibinfo {author} {\bibfnamefont {I.}~\bibnamefont {Liberal}}, \bibinfo {author} {\bibfnamefont {J.~E.}\ \bibnamefont {V{\'a}zquez-Lozano}},\ and\ \bibinfo {author} {\bibfnamefont {V.}~\bibnamefont {Pacheco-Pe{\~n}a}},\ }\bibfield  {title} {\bibinfo {title} {Quantum antireflection temporal coatings: quantum state frequency shifting and inhibited thermal noise amplification},\ }\href@noop {} {\bibfield  {journal} {\bibinfo  {journal} {Laser \& Photonics Reviews}\ }\textbf {\bibinfo {volume} {17}},\ \bibinfo {pages} {2200720} (\bibinfo {year} {2023})}\BibitemShut {NoStop}%
\bibitem [{\citenamefont {Ganfornina-Andrades}\ \emph {et~al.}(2023)\citenamefont {Ganfornina-Andrades}, \citenamefont {V{\'a}zquez-Lozano},\ and\ \citenamefont {Liberal}}]{ganfornina2023quantum}%
  \BibitemOpen
  \bibfield  {author} {\bibinfo {author} {\bibfnamefont {A.}~\bibnamefont {Ganfornina-Andrades}}, \bibinfo {author} {\bibfnamefont {J.~E.}\ \bibnamefont {V{\'a}zquez-Lozano}},\ and\ \bibinfo {author} {\bibfnamefont {I.}~\bibnamefont {Liberal}},\ }\bibfield  {title} {\bibinfo {title} {Quantum vacuum amplification in time-varying media with arbitrary temporal profiles},\ }\href@noop {} {\bibfield  {journal} {\bibinfo  {journal} {arXiv preprint arXiv:2312.13315}\ } (\bibinfo {year} {2023})}\BibitemShut {NoStop}%
\bibitem [{\citenamefont {Nation}\ \emph {et~al.}(2012)\citenamefont {Nation}, \citenamefont {Johansson}, \citenamefont {Blencowe},\ and\ \citenamefont {Nori}}]{nation2012colloquium}%
  \BibitemOpen
  \bibfield  {author} {\bibinfo {author} {\bibfnamefont {P.}~\bibnamefont {Nation}}, \bibinfo {author} {\bibfnamefont {J.}~\bibnamefont {Johansson}}, \bibinfo {author} {\bibfnamefont {M.}~\bibnamefont {Blencowe}},\ and\ \bibinfo {author} {\bibfnamefont {F.}~\bibnamefont {Nori}},\ }\bibfield  {title} {\bibinfo {title} {Colloquium: Stimulating uncertainty: Amplifying the quantum vacuum with superconducting circuits},\ }\href@noop {} {\bibfield  {journal} {\bibinfo  {journal} {Reviews of Modern Physics}\ }\textbf {\bibinfo {volume} {84}},\ \bibinfo {pages} {1} (\bibinfo {year} {2012})}\BibitemShut {NoStop}%
\bibitem [{\citenamefont {Wilson}\ \emph {et~al.}(2011)\citenamefont {Wilson}, \citenamefont {Johansson}, \citenamefont {Pourkabirian}, \citenamefont {Simoen}, \citenamefont {Johansson}, \citenamefont {Duty}, \citenamefont {Nori},\ and\ \citenamefont {Delsing}}]{wilson2011observation}%
  \BibitemOpen
  \bibfield  {author} {\bibinfo {author} {\bibfnamefont {C.~M.}\ \bibnamefont {Wilson}}, \bibinfo {author} {\bibfnamefont {G.}~\bibnamefont {Johansson}}, \bibinfo {author} {\bibfnamefont {A.}~\bibnamefont {Pourkabirian}}, \bibinfo {author} {\bibfnamefont {M.}~\bibnamefont {Simoen}}, \bibinfo {author} {\bibfnamefont {J.~R.}\ \bibnamefont {Johansson}}, \bibinfo {author} {\bibfnamefont {T.}~\bibnamefont {Duty}}, \bibinfo {author} {\bibfnamefont {F.}~\bibnamefont {Nori}},\ and\ \bibinfo {author} {\bibfnamefont {P.}~\bibnamefont {Delsing}},\ }\bibfield  {title} {\bibinfo {title} {Observation of the dynamical casimir effect in a superconducting circuit},\ }\href@noop {} {\bibfield  {journal} {\bibinfo  {journal} {nature}\ }\textbf {\bibinfo {volume} {479}},\ \bibinfo {pages} {376} (\bibinfo {year} {2011})}\BibitemShut {NoStop}%
\bibitem [{\citenamefont {L{\"a}hteenm{\"a}ki}\ \emph {et~al.}(2013)\citenamefont {L{\"a}hteenm{\"a}ki}, \citenamefont {Paraoanu}, \citenamefont {Hassel},\ and\ \citenamefont {Hakonen}}]{lahteenmaki2013dynamical}%
  \BibitemOpen
  \bibfield  {author} {\bibinfo {author} {\bibfnamefont {P.}~\bibnamefont {L{\"a}hteenm{\"a}ki}}, \bibinfo {author} {\bibfnamefont {G.}~\bibnamefont {Paraoanu}}, \bibinfo {author} {\bibfnamefont {J.}~\bibnamefont {Hassel}},\ and\ \bibinfo {author} {\bibfnamefont {P.~J.}\ \bibnamefont {Hakonen}},\ }\bibfield  {title} {\bibinfo {title} {Dynamical casimir effect in a josephson metamaterial},\ }\href@noop {} {\bibfield  {journal} {\bibinfo  {journal} {Proceedings of the National Academy of Sciences}\ }\textbf {\bibinfo {volume} {110}},\ \bibinfo {pages} {4234} (\bibinfo {year} {2013})}\BibitemShut {NoStop}%
\bibitem [{\citenamefont {Dodonov}(2020)}]{dodonov2020fifty}%
  \BibitemOpen
  \bibfield  {author} {\bibinfo {author} {\bibfnamefont {V.}~\bibnamefont {Dodonov}},\ }\bibfield  {title} {\bibinfo {title} {Fifty years of the dynamical casimir effect},\ }\href@noop {} {\bibfield  {journal} {\bibinfo  {journal} {Physics}\ }\textbf {\bibinfo {volume} {2}},\ \bibinfo {pages} {67} (\bibinfo {year} {2020})}\BibitemShut {NoStop}%
\bibitem [{sup()}]{suppmat}%
  \BibitemOpen
  \href@noop {} {}\bibinfo {note} {See Supplemental Material at [URL will be inserted by publisher] for detailed derivations of the equations found in the main text, as well as a discussion of the statistics of squeezed quantum fields.}\BibitemShut {Stop}%
\bibitem [{\citenamefont {Khurgin}(2023)}]{khurgin2023photonic}%
  \BibitemOpen
  \bibfield  {author} {\bibinfo {author} {\bibfnamefont {J.~B.}\ \bibnamefont {Khurgin}},\ }\bibfield  {title} {\bibinfo {title} {Photonic time crystals and parametric amplification: similarity and distinction},\ }\href@noop {} {\bibfield  {journal} {\bibinfo  {journal} {arXiv preprint arXiv:2305.15243}\ } (\bibinfo {year} {2023})}\BibitemShut {NoStop}%
\bibitem [{\citenamefont {Barnett}\ and\ \citenamefont {Radmore}(2002)}]{barnett2002methods}%
  \BibitemOpen
  \bibfield  {author} {\bibinfo {author} {\bibfnamefont {S.}~\bibnamefont {Barnett}}\ and\ \bibinfo {author} {\bibfnamefont {P.~M.}\ \bibnamefont {Radmore}},\ }\href@noop {} {\emph {\bibinfo {title} {Methods in theoretical quantum optics}}},\ Vol.~\bibinfo {volume} {15}\ (\bibinfo  {publisher} {Oxford University Press},\ \bibinfo {year} {2002})\BibitemShut {NoStop}%
\bibitem [{\citenamefont {Milburn}(2012)}]{milburn2012quantum}%
  \BibitemOpen
  \bibfield  {author} {\bibinfo {author} {\bibfnamefont {G.}~\bibnamefont {Milburn}},\ }\bibfield  {title} {\bibinfo {title} {Quantum optics},\ }\href@noop {} {\bibfield  {journal} {\bibinfo  {journal} {Springer Handbook of Lasers and Optics}\ ,\ \bibinfo {pages} {1305}} (\bibinfo {year} {2012})}\BibitemShut {NoStop}%
\bibitem [{\citenamefont {Gerry}\ and\ \citenamefont {Knight}(2023)}]{gerry2023introductory}%
  \BibitemOpen
  \bibfield  {author} {\bibinfo {author} {\bibfnamefont {C.~C.}\ \bibnamefont {Gerry}}\ and\ \bibinfo {author} {\bibfnamefont {P.~L.}\ \bibnamefont {Knight}},\ }\href@noop {} {\emph {\bibinfo {title} {Introductory quantum optics}}}\ (\bibinfo  {publisher} {Cambridge university press},\ \bibinfo {year} {2023})\BibitemShut {NoStop}%
\bibitem [{\citenamefont {Park}\ \emph {et~al.}(2022)\citenamefont {Park}, \citenamefont {Park}, \citenamefont {Lee}, \citenamefont {Shin}, \citenamefont {Ryu}, \citenamefont {Jeon}, \citenamefont {Park},\ and\ \citenamefont {Min}}]{park2022comment}%
  \BibitemOpen
  \bibfield  {author} {\bibinfo {author} {\bibfnamefont {J.}~\bibnamefont {Park}}, \bibinfo {author} {\bibfnamefont {H.~C.}\ \bibnamefont {Park}}, \bibinfo {author} {\bibfnamefont {K.}~\bibnamefont {Lee}}, \bibinfo {author} {\bibfnamefont {J.}~\bibnamefont {Shin}}, \bibinfo {author} {\bibfnamefont {J.-W.}\ \bibnamefont {Ryu}}, \bibinfo {author} {\bibfnamefont {W.}~\bibnamefont {Jeon}}, \bibinfo {author} {\bibfnamefont {N.}~\bibnamefont {Park}},\ and\ \bibinfo {author} {\bibfnamefont {B.}~\bibnamefont {Min}},\ }\bibfield  {title} {\bibinfo {title} {Comment on" amplified emission and lasing in photonic time crystals"},\ }\href@noop {} {\bibfield  {journal} {\bibinfo  {journal} {arXiv preprint arXiv:2211.14832}\ } (\bibinfo {year} {2022})}\BibitemShut {NoStop}%
\bibitem [{\citenamefont {Ashida}\ \emph {et~al.}(2020)\citenamefont {Ashida}, \citenamefont {Gong},\ and\ \citenamefont {Ueda}}]{ashida2020non}%
  \BibitemOpen
  \bibfield  {author} {\bibinfo {author} {\bibfnamefont {Y.}~\bibnamefont {Ashida}}, \bibinfo {author} {\bibfnamefont {Z.}~\bibnamefont {Gong}},\ and\ \bibinfo {author} {\bibfnamefont {M.}~\bibnamefont {Ueda}},\ }\bibfield  {title} {\bibinfo {title} {Non-hermitian physics},\ }\href@noop {} {\bibfield  {journal} {\bibinfo  {journal} {Advances in Physics}\ }\textbf {\bibinfo {volume} {69}},\ \bibinfo {pages} {249} (\bibinfo {year} {2020})}\BibitemShut {NoStop}%
\bibitem [{\citenamefont {Boyd}\ \emph {et~al.}(2008)\citenamefont {Boyd}, \citenamefont {Gaeta},\ and\ \citenamefont {Giese}}]{boyd2008nonlinear}%
  \BibitemOpen
  \bibfield  {author} {\bibinfo {author} {\bibfnamefont {R.~W.}\ \bibnamefont {Boyd}}, \bibinfo {author} {\bibfnamefont {A.~L.}\ \bibnamefont {Gaeta}},\ and\ \bibinfo {author} {\bibfnamefont {E.}~\bibnamefont {Giese}},\ }\bibfield  {title} {\bibinfo {title} {Nonlinear optics},\ }in\ \href@noop {} {\emph {\bibinfo {booktitle} {Springer Handbook of Atomic, Molecular, and Optical Physics}}}\ (\bibinfo  {publisher} {Springer},\ \bibinfo {year} {2008})\ pp.\ \bibinfo {pages} {1097--1110}\BibitemShut {NoStop}%
\bibitem [{\citenamefont {Drummond}\ and\ \citenamefont {Hillery}(2014)}]{drummond2014quantum}%
  \BibitemOpen
  \bibfield  {author} {\bibinfo {author} {\bibfnamefont {P.~D.}\ \bibnamefont {Drummond}}\ and\ \bibinfo {author} {\bibfnamefont {M.}~\bibnamefont {Hillery}},\ }\href@noop {} {\emph {\bibinfo {title} {The quantum theory of nonlinear optics}}}\ (\bibinfo  {publisher} {Cambridge University Press},\ \bibinfo {year} {2014})\BibitemShut {NoStop}%
\bibitem [{\citenamefont {Glauber}(1963)}]{glauber1963quantum}%
  \BibitemOpen
  \bibfield  {author} {\bibinfo {author} {\bibfnamefont {R.~J.}\ \bibnamefont {Glauber}},\ }\bibfield  {title} {\bibinfo {title} {The quantum theory of optical coherence},\ }\href@noop {} {\bibfield  {journal} {\bibinfo  {journal} {Physical Review}\ }\textbf {\bibinfo {volume} {130}},\ \bibinfo {pages} {2529} (\bibinfo {year} {1963})}\BibitemShut {NoStop}%
\bibitem [{\citenamefont {Kim}\ \emph {et~al.}(1989)\citenamefont {Kim}, \citenamefont {De~Oliveira},\ and\ \citenamefont {Knight}}]{kim1989properties}%
  \BibitemOpen
  \bibfield  {author} {\bibinfo {author} {\bibfnamefont {M.}~\bibnamefont {Kim}}, \bibinfo {author} {\bibfnamefont {F.}~\bibnamefont {De~Oliveira}},\ and\ \bibinfo {author} {\bibfnamefont {P.}~\bibnamefont {Knight}},\ }\bibfield  {title} {\bibinfo {title} {Properties of squeezed number states and squeezed thermal states},\ }\href@noop {} {\bibfield  {journal} {\bibinfo  {journal} {Physical Review A}\ }\textbf {\bibinfo {volume} {40}},\ \bibinfo {pages} {2494} (\bibinfo {year} {1989})}\BibitemShut {NoStop}%
\bibitem [{\citenamefont {Dikopoltsev}\ \emph {et~al.}(2022)\citenamefont {Dikopoltsev}, \citenamefont {Sharabi}, \citenamefont {Lyubarov}, \citenamefont {Lumer}, \citenamefont {Tsesses}, \citenamefont {Lustig}, \citenamefont {Kaminer},\ and\ \citenamefont {Segev}}]{dikopoltsev2022light}%
  \BibitemOpen
  \bibfield  {author} {\bibinfo {author} {\bibfnamefont {A.}~\bibnamefont {Dikopoltsev}}, \bibinfo {author} {\bibfnamefont {Y.}~\bibnamefont {Sharabi}}, \bibinfo {author} {\bibfnamefont {M.}~\bibnamefont {Lyubarov}}, \bibinfo {author} {\bibfnamefont {Y.}~\bibnamefont {Lumer}}, \bibinfo {author} {\bibfnamefont {S.}~\bibnamefont {Tsesses}}, \bibinfo {author} {\bibfnamefont {E.}~\bibnamefont {Lustig}}, \bibinfo {author} {\bibfnamefont {I.}~\bibnamefont {Kaminer}},\ and\ \bibinfo {author} {\bibfnamefont {M.}~\bibnamefont {Segev}},\ }\bibfield  {title} {\bibinfo {title} {Light emission by free electrons in photonic time-crystals},\ }\href@noop {} {\bibfield  {journal} {\bibinfo  {journal} {Proceedings of the National Academy of Sciences}\ }\textbf {\bibinfo {volume} {119}},\ \bibinfo {pages} {e2119705119} (\bibinfo {year} {2022})}\BibitemShut {NoStop}%
\end{thebibliography}%


\begin{thebibliography}{1}

\bibitem{ashida2020non}
Yuto Ashida, Zongping Gong, and Masahito Ueda.
\newblock Non-hermitian physics.
\newblock {\em Advances in Physics}, 69(3):249--435, 2020.

\bibitem{mendoncca2003temporal}
JT~Mendon{\c{c}}a, AM~Martins, and A~Guerreiro.
\newblock Temporal beam splitter and temporal interference.
\newblock {\em Physical Review A}, 68(4):043801, 2003.

\bibitem{barnett2002methods}
Stephen Barnett and Paul~M Radmore.
\newblock {\em Methods in theoretical quantum optics}, volume~15.
\newblock Oxford University Press, 2002.

\bibitem{gerry2023introductory}
Christopher~C Gerry and Peter~L Knight.
\newblock {\em Introductory quantum optics}.
\newblock Cambridge university press, 2023.

\bibitem{glauber1963quantum}
Roy~J Glauber.
\newblock The quantum theory of optical coherence.
\newblock {\em Physical Review}, 130(6):2529, 1963.

\bibitem{boyd2008nonlinear}
Robert~W Boyd, Alexander~L Gaeta, and Enno Giese.
\newblock Nonlinear optics.
\newblock In {\em Springer Handbook of Atomic, Molecular, and Optical Physics}, pages 1097--1110. Springer, 2008.

\bibitem{drummond2014quantum}
Peter~D Drummond and Mark Hillery.
\newblock {\em The quantum theory of nonlinear optics}.
\newblock Cambridge University Press, 2014.

\bibitem{kim1989properties}
MS~Kim, FAM De~Oliveira, and PL~Knight.
\newblock Properties of squeezed number states and squeezed thermal states.
\newblock {\em Physical Review A}, 40(5):2494, 1989.

\end{thebibliography}

\end{document}


\maketitle

\section*{Transfer matrix and time evolution}

We start from Maxwell curl equations in a source-free magnetodielectric medium:

\begin{subequations}
    \begin{equation}
        \curl\boldsymbol{E} = -\frac{\partial\boldsymbol{B}}{\partial t}
        \label{eq:faraday}
    \end{equation}
    and
    \begin{equation}
       \curl\boldsymbol{H} = \frac{\partial\boldsymbol{D}}{\partial t},
       \label{eq:ampere}
    \end{equation}
\end{subequations}

which we supplement with the following constitutive relations:

\begin{subequations}
    \begin{equation}
        \boldsymbol{D} = \varepsilon(t)\boldsymbol{E}
        \label{eq:d_e_relation}
    \end{equation}
    and
    \begin{equation}
        \boldsymbol{B} = \mu(t)\boldsymbol{H},
        \label{eq:b_h_relation}
    \end{equation}
\end{subequations}

where we assume a linear, non-dispersive (and hence, local in time) but time-modulated response.\\
By taking time derivatives in Faraday and Ampère laws \ref{eq:faraday} and \ref{eq:ampere}, and making use of the constitutive relations \ref{eq:d_e_relation} and \ref{eq:b_h_relation}, we arrive to the following wave equations for the displacement field and the magnetic induction:

\begin{subequations}
    \begin{equation}
        \curl\curl\boldsymbol{D} + \varepsilon(t)\frac{\partial}{\partial t}\left(\mu(t)\frac{\partial\boldsymbol{D}}{\partial t}\right) = \boldsymbol{0},
        \label{eq:wave_eq_d}
    \end{equation}
    and
    \begin{equation}
        \curl\curl\boldsymbol{B} + \mu(t)\frac{\partial}{\partial t}\left(\varepsilon(t)\frac{\partial\boldsymbol{B}}{\partial t}\right) = \boldsymbol{0}.
        \label{eq:wave_eq_b}
    \end{equation}
\end{subequations}

Now going to Fourier space, for given $\boldsymbol{k}$ we have two distinct polarizations $\boldsymbol{e}_{\boldsymbol{k}\sigma}\text{ }(\sigma = 1,2)$, with the $\sigma$-polarized fields solving the differential equations found below:

\begin{subequations}
    \begin{equation}
         \varepsilon(t)\frac{\partial}{\partial t}\left(\mu(t)\frac{\partial D_{\boldsymbol{k}\sigma}}{\partial t}\right) -k^2D_{\boldsymbol{k}\sigma} = 0,
        \label{eq:wave_eq_d_fourier}
    \end{equation}
    and
    \begin{equation}
        \mu(t)\frac{\partial}{\partial t}\left(\varepsilon(t)\frac{\partial B_{\boldsymbol{k}\sigma}}{\partial t}\right) -k^2B_{\boldsymbol{k}\sigma} = 0.
        \label{eq:wave_eq_b_fourier}
    \end{equation}
\end{subequations}

A glance at the wave equations shows that neither of the time operators acting on $\boldsymbol{D}$ and $\boldsymbol{B}$ is Hermitian (it suffices to take the adjoint of each one of them to see this); what's more, the time operator acting on $\boldsymbol{D}$ is the adjoint of the one acting on $\boldsymbol{B}$. Actually, this lack of Hermiticity, rooted in an impedance-mismatch, shall be responsible for the amplification of waves in time-varying media. However, when the impedance $Z(t) = \sqrt{\mu(t)/\varepsilon(t)}$ is set to a constant value, backscattering disappears and both operators become self-adjoint and furthermore, equal to each other.\\
From the wave equations we can infer the temporal boundary conditions for the fields

\begin{subequations}
    \begin{equation}
        \mu(t_+)\frac{\partial\boldsymbol{D}(t_+)}{\partial t} = \mu(t_-)\frac{\partial\boldsymbol{D}(t_-)}{\partial t}
        \label{eq:boundary_condition_d}
    \end{equation}
    and
    \begin{equation}
        \varepsilon(t_+)\frac{\partial\boldsymbol{B}(t_+)}{\partial t} = \varepsilon(t_-)\frac{\partial\boldsymbol{B}(t_-)}{\partial t},
        \label{eq:boundary_condition_b}
    \end{equation}
\end{subequations}

where $t_+$ and $t_-$ are infinitely close to the instant $t$ in which the optical properties of the system experience a sudden and discontinuous change in their values.
\par
From now on we will take $\mu(t) = \mu_1$ and focus on the case in which only the permittivity is time-varying, thus making the displacement field and its first time derivative continuous at all times. We assume $\varepsilon(t)$ varies in a step-like manner, alternating between two values $\varepsilon_a$ and $\varepsilon_b$, with each slab of constant permittivity lasting for a time $t_a$ and $t_b$, respectively. What's more, we write $\varepsilon(t) = \varepsilon_0n^2(t)$ and work with the refractive index.\\
In a constant refractive index time interval, solutions to Maxwell equations are simply those of free-space with the appropriate re-scaling of the optical properties. Thus, solutions to \ref{eq:wave_eq_b_fourier} within the $N$-th $n(t) = n_a$ temporal slab may be written as

\begin{equation}
    B^{(N)}_{\boldsymbol{k}\sigma}(t) = B^{(N)}_{\boldsymbol{k}\sigma,+}e^{-i\frac{\omega_k}{n_a}t} + B^{(N)}_{\boldsymbol{k}\sigma,-}e^{i\frac{\omega_k}{n_a}t},
\end{equation}

where both positive and negative frequencies have been accounted for.\\
The total field is then

\begin{equation}
    \boldsymbol{B}^{(N)}(\boldsymbol{x},t) = \int\frac{d^3k}{(2\pi)^3}e^{i\boldsymbol{k}\cdot\boldsymbol{x}}\sum_\sigma \Big(B^{(N)}_{\boldsymbol{k}\sigma,+}e^{-i\frac{\omega_k}{n_a}t} + B^{(N)}_{\boldsymbol{k}\sigma,-}e^{i\frac{\omega_k}{n_a}t}\Big)\boldsymbol{e}_{\boldsymbol{k}\sigma},
\end{equation}

which needs to be real. From the reality constraint and the fact that $\boldsymbol{e}_{-\boldsymbol{k}\sigma} = (-1)^{\sigma-1}\boldsymbol{e}_{\boldsymbol{k}\sigma}$ (the $\sigma=2$ polarization flips its direction when the wavevector is inverted), we arrive to $B^{(N)}_{\boldsymbol{k}\sigma,-} = (-1)^{\sigma-1}(B^{(N)}_{-\boldsymbol{k}\sigma,+})^*$. Thus, we see the negative frequency terms are just the complex conjugates of the positive frequency ones of the corresponding backward waves (save for an additional $\pi$-shift for the case of $\sigma = 2$ polarization).\\
We now derive the transfer matrix of our photonic time crystal. We have two slabs, one with $n(t)=n_a$ lasting for a time $t_a$ followed by a $n(t)=n_b$ one, which lasts for a time $t_b$, with $T = t_a + t_b$ being the period of our temporal modulation. The magnetic induction flux is continuous at time interfaces, but exhibits a finite jump discontinuity in its first time derivative, as seen in equation \ref{eq:boundary_condition_b}. Therefore, at a time $t = NT + t_a$ and for the $N$-th period interface between $n_a$ and $n_b$-type slabs, we have the following two equations: 

\begin{equation}
    B^{(N)}_{\boldsymbol{k}\sigma,b+}e^{-i\frac{\omega_k}{n_b}(NT+t_a)} + B^{(N)}_{\boldsymbol{k}\sigma,b-}e^{i\frac{\omega_k}{n_b}(NT+t_a)} = B^{(N)}_{\boldsymbol{k}\sigma,a+}e^{-i\frac{\omega_k}{n_a}(NT+t_a)} + B^{(N)}_{\boldsymbol{k}\sigma,a-}e^{i\frac{\omega_k}{n_a}(NT+t_a)}
\end{equation}

and

\begin{equation}
    -\frac{1}{n_b}B^{(N)}_{\boldsymbol{k}\sigma,b+}e^{-i\frac{\omega_k}{n_b}(NT+t_a)} + \frac{1}{n_b}B^{(N)}_{\boldsymbol{k}\sigma,b-}e^{i\frac{\omega_k}{n_b}(NT+t_a)} = -\frac{1}{n_a}B^{(N)}_{\boldsymbol{k}\sigma,a+}e^{-i\frac{\omega_k}{n_a}(NT+t_a)} + \frac{1}{n_a}B^{(N)}_{\boldsymbol{k}\sigma,a-}e^{i\frac{\omega_k}{n_a}(NT+t_a)},
\end{equation}

where we have added the $a$ and $b$ subindices specifying the value taken by the refractive index.\\
To derive a recursive formula connecting the fields of the $n(t)=n_a$ slabs, we need to consider the second type of interface encountered in this system, the one between the $N$-th period $n_b$-type slab and the $N+1$-th period $n_a$-type one, which happens at $t = (N+1)T$. This second interface leads to the a second pair of equations:

\begin{equation}
    B^{(N+1)}_{\boldsymbol{k}\sigma,a+}e^{-i\frac{\omega_k}{n_a}(N+1)T} + B^{(N+1)}_{\boldsymbol{k}\sigma,a-}e^{i\frac{\omega_k}{n_a}(N+1)T} = B^{(N)}_{\boldsymbol{k}\sigma,b+}e^{-i\frac{\omega_k}{n_b}(N+1)T} + B^{(N)}_{\boldsymbol{k}\sigma,b-}e^{i\frac{\omega_k}{n_b}(N+1)T}
\end{equation}

and

\begin{equation}
    -\frac{1}{n_a}B^{(N+1)}_{\boldsymbol{k}\sigma,a+}e^{-i\frac{\omega_k}{n_a}(N+1)T} +\frac{1}{n_a}B^{(N+1)}_{\boldsymbol{k}\sigma,a-}e^{i\frac{\omega_k}{n_a}(N+1)T} = -\frac{1}{n_b}B^{(N)}_{\boldsymbol{k}\sigma,b+}e^{-i\frac{\omega_k}{n_b}(N+1)T} +\frac{1}{n_b}B^{(N)}_{\boldsymbol{k}\sigma,b-}e^{i\frac{\omega_k}{n_b}(N+1)T}.
\end{equation}

In order to further simplify the problem, we redefine the field amplitudes as $B^{(N)}_{\boldsymbol{k}\sigma,a\pm} = \Tilde{B}^{(N)}_{\boldsymbol{k}\sigma,a\pm}\exp(\pm i\omega_k(NT+t_a)/n_a)$ and $B^{(N)}_{\boldsymbol{k}\sigma,b\pm} = \Tilde{B}^{(N)}_{\boldsymbol{k}\sigma,b\pm}\exp(\pm i\omega_k(NT+t_a)/n_b)$. This will make the transfer matrix for the $\tilde{B}$ amplitudes independent of the period of the PTC; on the other hand, since it is only a change of phase, it is a canonical transformation.\\
If we combine the two sets of equations, we arrive to the following transfer matrix recursive formula for the magnetic induction flux:

\begin{equation}
\begin{pmatrix}
    \Tilde{B}_{\boldsymbol{k}\sigma,a-}^{(N+1)} \\ \\\Tilde{B}_{\boldsymbol{k}\sigma,a+}^{(N+1)}
\end{pmatrix} = \boldsymbol{T}\cdot \begin{pmatrix}
    \Tilde{B}_{\boldsymbol{k}\sigma,a-}^{(N)} \\ \\ \Tilde{B}_{\boldsymbol{k}\sigma,a+}^{(N)}
\end{pmatrix},
\end{equation}

where the transfer matrix $\boldsymbol{T}$ reads

\begin{equation}
    \boldsymbol{T} = \begin{pmatrix}
        \frac{e^{i\frac{\omega_kt_a}{n_a}}}{n_an_b}\Big(n_an_b\cos(\frac{\omega_kt_b}{n_b}) + i\frac{n_b^2+n_a^2}{2}\sin(\frac{\omega_kt_b}{n_b})\Big) & i\frac{e^{i\frac{\omega_kt_a}{n_a}}}{2n_an_b}(n_b^2-n_a^2)\sin(\frac{\omega_kt_b}{n_b}) \\
    -i\frac{e^{-i\frac{\omega_kt_a}{n_a}}}{2n_an_b}(n_b^2-n_a^2)\sin(\frac{\omega_kt_b}{n_b}) & \frac{e^{-i\frac{\omega_kt_a}{n_a}}}{n_an_b}\Big(n_an_b\cos(\frac{\omega_kt_b}{n_b}) - i\frac{n_b^2+n_a^2}{2}\sin(\frac{\omega_kt_b}{n_b})\Big)
    \end{pmatrix}.
    \label{eq:transfer_matrix}
\end{equation}

We mention here that the transfer matrix is not unique owing to the freedom we have to change the phase of the fields in a conjugated manner. However, since the latter amounts for a diagonal unitary transformation, it does not alter the spectrum of the transfer matrix. Also, all the properties of the transfer matrix mentioned in the main text can be verified: $(\boldsymbol{T}^N)_{11} = \big((\boldsymbol{T}^N)_{22}\big)^*$ and $(\boldsymbol{T}^N)_{12} = \big((\boldsymbol{T}^N)_{21}\big)^*$, as well as $\det(\boldsymbol{T}) = 1$\\
The transfer matrix has eigenvalues given by $\lambda_{\pm} = p\text{ }\mp\text{ }i\sqrt{1-p^2}$, where $p = \cos(\frac{\omega_kt_a}{n_a})\cos(\frac{\omega_kt_b}{n_b})-\frac{n_a^2+n_b^2}{2n_an_b}\sin(\frac{\omega_kt_a}{n_a})\sin(\frac{\omega_kt_b}{n_b})$. We see $p$ controls whether the eigenvalues are complex-valued with moduli unity and complex conjugates between themselves, which happens for $|p|<1$, or if, on the other hand, they are real-valued, which occurs when $|p|>1$; in both cases we have $\lambda_+\lambda_- = 1$. For the $p = 1$ case, the transfer matrix cannot be diagonalized and we have an exceptional point \cite{ashida2020non}. However, the general formulae derived for the diagonalizable case can still be applied to the non-diagonalizable one; this is equivalent to being outside the exceptional point, but infinitely close to it, with $p = 1 + \delta$, and taking the limit $\delta\longrightarrow0$ at the very end of the calculations.\\
Lastly, the $\lambda_+\lambda_- = 1$ constraint enables us to introduce a Floquet frequency through $\lambda_{\pm} = \exp(\mp i\omega_FT)$, where $\omega_F = i\ln(p-i\sqrt{1-p^2})/T$, which is defined modulo $2\pi/T$ (in the entirety of the main text, we concentrate on the first Brillouin zone).\\
Since we shall be concerned with what happens at the $n(t)=n_a$ time intervals, we can drop the $a$ and $b$ subscripts. Furthermore, owing to the relationship between negative frequencies and backward waves, we can drop the frequency sign subscript and work with positive frequency amplitudes and their complex conjugates, thus writing

\begin{equation*}
\begin{pmatrix}
(-1)^{\lambda-1}(\Tilde{B}_{-\boldsymbol{k}\sigma}^{(N+1)})^* \\ \\
    \Tilde{B}_{\boldsymbol{k}\sigma}^{(N+1)}
\end{pmatrix} = \boldsymbol{T}\cdot \begin{pmatrix} (-1)^{\lambda-1}(\Tilde{B}_{-\boldsymbol{k}\sigma}^{(N)})^* \\ \\
    \Tilde{B}_{\boldsymbol{k}\sigma}^{(N)}
\end{pmatrix}.
\end{equation*}

The $(-1)^{\sigma-1}$ multiplying the backward wave amplitude can be factored out as $-\boldsymbol{\sigma}_z$, with $\boldsymbol{\sigma}_z = \text{diag}(1,-1)$, the corresponding Pauli matrix. Thus, we can write the recursive formula as

\begin{equation*}
\begin{pmatrix}
(\Tilde{B}_{-\boldsymbol{k}\sigma}^{(N+1)})^* \\ \\
    \Tilde{B}_{\boldsymbol{k}\sigma}^{(N+1)}
\end{pmatrix} = (\boldsymbol{\sigma}_z^{\sigma-1}\cdot\boldsymbol{T}\cdot\boldsymbol{\sigma}_z^{\sigma-1}) \begin{pmatrix} (\Tilde{B}_{-\boldsymbol{k}\sigma}^{(N)})^* \\ \\
    \Tilde{B}_{\boldsymbol{k}\sigma}^{(N)}
\end{pmatrix}.
\end{equation*}

It is straightforward to prove that $\boldsymbol{\sigma}_z^{\sigma-1}\cdot\boldsymbol{T}\cdot\boldsymbol{\sigma}_z^{\sigma-1}$ coincides with $\boldsymbol{T}$ except for the off-diagonal elements, which are multiplied by $(-1)^{\sigma-1}$. Photon pair creation wise, this amounts to $\pi$-shifting the squeezing phase $\varphi$ for the $\sigma=2-$polarized photons, with $\varphi\longrightarrow\varphi+\pi$. Since the squeezing phase $\varphi$ shall not be important for calculating photon pair creation probabilities, we need not worry about the polarization of the mode. Hence, if we further drop the $\sim$ on top of the amplitudes, we can write the transfer matrix recursive formula like

\begin{equation}
\begin{pmatrix}
(B_{-\boldsymbol{k}\sigma}^{(N+1)})^* \\ \\
    B_{\boldsymbol{k}\sigma}^{(N+1)}
\end{pmatrix} = \boldsymbol{T}\cdot\begin{pmatrix} (B_{-\boldsymbol{k}\sigma}^{(N)})^* \\ \\
    B_{\boldsymbol{k}\sigma}^{(N)}
\end{pmatrix},
\end{equation}

which leads to

\begin{equation}
\begin{pmatrix}
(B_{-\boldsymbol{k}\sigma}^{(N)})^* \\ \\
    B_{\boldsymbol{k}\sigma}^{(N)}
\end{pmatrix} = \boldsymbol{T}^N\cdot\begin{pmatrix} (B_{-\boldsymbol{k}\sigma}^{(0)})^* \\ \\
    B_{\boldsymbol{k}\sigma}^{(0)}
\end{pmatrix},
\end{equation}

the one used in the main text.\\
Lastly, we derive the expression for $(\boldsymbol{T}^N)_{12}$ encountered in the manuscript. We begin by diagonalizing the transfer matrix: $\boldsymbol{T} = \boldsymbol{S}\cdot\text{diag}(\lambda_-,\lambda_+)\cdot(\boldsymbol{S})^{-1}$, where the diagonalization matrix $\boldsymbol{S}$ (which, like $\boldsymbol{T}$, is not uniquely defined) is given by

\begin{equation*}
    \boldsymbol{S} = \begin{pmatrix}
        1 & -\frac{T_{12}}{T_{11}-\lambda_+} \\ -\frac{T_{21}}{T_{22}-\lambda_-} & 1
    \end{pmatrix}.
\end{equation*}

It is straightforward to prove that $\boldsymbol{T}^N = \boldsymbol{S}\cdot\text{diag}(\lambda_-^N,\lambda_+^N)\cdot(\boldsymbol{S})^{-1}$. After inverting $\boldsymbol{S}$ and performing the matrix calculations explicitly, we arrive to $(\boldsymbol{T}^N)_{12} = -S_{11}S_{12}(\lambda_-^N-\lambda_+^N)/|\boldsymbol{S}|$, which we can further write as $(\boldsymbol{T}^N)_{12} = -i2S_{11}S_{12} \sin(N\omega_FT)/|\boldsymbol{S}|$. By writing $S_{12}$ explicitly, we have $(\boldsymbol{T}^N)_{12} = i2(T_{12}/(T_{11}-\lambda_+))\sin(N\omega_FT)/|\boldsymbol{S}|$. Since for $N = 1$ we need the latter to be just $T_{12} = (\boldsymbol{T})_{12}$, we prove that $\sin(\omega_FT) = |\boldsymbol{S}|(T_{11}-\lambda_+)/i2$, which leads us to the desired formula:

\begin{equation}
    (\boldsymbol{T}^N)_{12} = \frac{\sin(N\omega_FT)}{\sin(\omega_FT)}(\boldsymbol{T})_{12}.
    \label{eq:T_12_N_slabs}
\end{equation}

Even though we have derived the above on the assumption of diagonalizability of the transfer matrix, it is well-defined for $\omega_F\longrightarrow0$, which happens at the exceptional point. On the other hand, its zeroes are given by $\omega_F = n\pi/NT$, where $n = 1, 2, ...,N-1$, as well as $T_{12} = 0$; the latter can be easily calculated by looking at equation \ref{eq:transfer_matrix}, and come from $\omega_kt_b/n_b = n\pi$, with $n$ a positive integer (including zero). Thus, we have $N$ different transparency conditions for $N$ applications of the transfer matrix, each of which yields infinitely many modes for which the reflectivity vanishes. We also mention here that the transparency resonances happen for real Floquet frequency and hence, outside of the momentum bandgaps, as one would intuitively expect.\\
There is a second comment pending: equation \ref{eq:T_12_N_slabs} also applies to the case of a continuous (and periodic) modulation, since it only assumes that $\det(\boldsymbol{T}) = 1$. The latter property was used to parameterize the spectra of the $\boldsymbol{T}$-matrix in terms of a Floquet frequency through $\lambda_{\pm} = \exp(\mp i\omega_FT)$. Hence, even though the exact form of the Floquet frequency $\omega_F$ may differ from that of a step-like modulation, the phenomenology of the PTC remains the same: modes within the bandgaps (complex Floquet frequencies) are exponentially amplified, while those inside the bands progressively become transparent to the time modulation.

\section*{Photon pair creation}

To quantize the field, we shall work with the vector potential. The latter and the magnetic induction flux are related in Fourier space by $\boldsymbol{B}_{\boldsymbol{k}} = i\boldsymbol{k}\cross\boldsymbol{A}_{\boldsymbol{k}}$ and so, they share the same transfer matrix\footnote{They do not share it with the displacement field, though, whose transfer matrix has $\pi-$shifted its off-diagonal elements with respect to to theirs.}.\\
We want the vector potential and the canonical conjugate momentum, given by $\boldsymbol{\Pi} = -\boldsymbol{D}$, to satisfy the equal times canonical commutation relations $[\boldsymbol{\Hat{A}}(\boldsymbol{x},t),\boldsymbol{\Hat{\Pi}}(\boldsymbol{x}',t)] = i\hbar\delta(\boldsymbol{x}-\boldsymbol{x}')\boldsymbol{I}$. To do so, we ought to have $[\Hat{A}^{(N)}_{\boldsymbol{k}\sigma},(\Hat{\Pi}^{(N)}_{\boldsymbol{k}'\sigma'})^\dagger] = (i\hbar/2)(2\pi)^3\delta(\boldsymbol{k}-\boldsymbol{k}')\delta_{\sigma,\sigma'}$ satisfied too. Since $\boldsymbol{D}=\varepsilon(t)\boldsymbol{E} = -\varepsilon(t)\partial_t\boldsymbol{A}$, we actually have $\Hat{\Pi}^{(N)}_{\boldsymbol{k}\sigma} = -i\omega_kn_a\Hat{A}^{(N)}_{\boldsymbol{k}\sigma}$ within the $n(t) = n_a$ time intervals. As a consequence, we have $[\Hat{A}^{(N)}_{\boldsymbol{k}\sigma},(\Hat{A}^{(N)}_{\boldsymbol{k}'\sigma'})^\dagger] = (\hbar/2\omega_kn_a)(2\pi)^3\delta(\boldsymbol{k}-\boldsymbol{k}')\delta_{\sigma,\sigma'}$ and we introduce the creation and annihilation operators as $(\Hat{A}^{(N)}_{\boldsymbol{k}\sigma})^\dagger = \sqrt{\hbar/2\omega_kn_a}(\hat{a}^{(N)}_{\boldsymbol{k}\sigma})^\dagger$ and $\Hat{A}^{(N)}_{\boldsymbol{k}\sigma} = \sqrt{\hbar/2\omega_kn_a}\hat{a}^{(N)}_{\boldsymbol{k}\sigma}$, which satisfy $[\hat{a}^{(N)}_{\boldsymbol{k}\sigma},(\hat{a}^{(N)}_{\boldsymbol{k}'\sigma'})^\dagger] = (2\pi)^3\delta(\boldsymbol{k}-\boldsymbol{k}')\delta_{\sigma,\sigma'}$.
\par
Now that we have quantized the field, we can connect the creation and annihilation operators of different slabs using the transfer matrix:

\begin{equation}
\begin{pmatrix}
(\hat{a}_{-\boldsymbol{k}}^{(N)})^\dagger \\ \\
    \hat{a}_{\boldsymbol{k}}^{(N)}
\end{pmatrix} = \boldsymbol{T}^N\cdot\begin{pmatrix} (\hat{a}_{-\boldsymbol{k}}^{(0)})^\dagger \\ \\
    \hat{a}_{\boldsymbol{k}}^{(0)}
\end{pmatrix},
\end{equation}

where as in the manuscript, we have dropped the polarization indices because, as seen in the previous section, the transfer matrix does not couple photons whose polarizations differ.\\
Since $\det(\boldsymbol{T}) = 1$, the transfer matrix indeed induces a canonical transformation between the Fock space operators of different slabs. This, together with the other properties of the transfer matrix, enables us to cast the following Bogoliubov transformation \cite{mendoncca2003temporal}:

\begin{subequations}
\begin{equation}
    (\Hat{a}_{-\boldsymbol{k}}^{(N)})^\dagger = \cosh(r)e^{i\theta_1}(\Hat{a}_{-\boldsymbol{k}}^{(0)})^\dagger + \sinh(r)e^{i\theta_2}\hat{a}^{(0)}_{\boldsymbol{k}},
    \label{eq:bogoliubov_1}
\end{equation}
\begin{equation}
    \hat{a}^{(N)}_{\boldsymbol{k}} = \cosh(r)e^{-i\theta_1}\hat{a}^{(0)}_{\boldsymbol{k}} + \sinh(r)e^{-i\theta_2}(\Hat{a}_{-\boldsymbol{k}}^{(0)})^\dagger.
    \label{eq:bogoliubov_2}
\end{equation}
\end{subequations}

If we now multiply the above pair of equations by $\exp(\mp i\theta_1)$, respectively, we get:

\begin{subequations}
\begin{equation*}
    (\Hat{a}_{-\boldsymbol{k}}^{(N)})^\dagger e^{-i\theta_1} = \cosh(r)(\Hat{a}_{-\boldsymbol{k}}^{(0)})^\dagger + \sinh(r)e^{-i(\theta_1 -\theta_2)}\hat{a}^{(0)}_{\boldsymbol{k}},
    \label{eq:bogoliubov_1}
\end{equation*}
\begin{equation*}
    \hat{a}^{(N)}_{\boldsymbol{k}}e^{i\theta_1} = \cosh(r)\hat{a}^{(0)}_{\boldsymbol{k}} + \sinh(r)e^{i(\theta_1 -\theta_2)}(\Hat{a}_{-\boldsymbol{k}}^{(0)})^\dagger.
    \label{eq:bogoliubov_2}
\end{equation*}
\end{subequations}

We can now redefine $\hat{a}^{(N)}_{\boldsymbol{k}}\exp(i\theta_1)\longrightarrow\hat{a}^{(N)}_{\boldsymbol{k}}$ and $(\hat{a}^{(N)}_{-\boldsymbol{k}})^\dagger\exp(-i\theta_1)\longrightarrow(\hat{a}^{(N)}_{-\boldsymbol{k}})^\dagger$, which, upon introduction of the relative phase between forward and backward waves $\varphi = \theta_1 - \theta_2$, allows us to write the Bogoliubov transformation as

\begin{subequations}
\begin{equation}
    (\Hat{a}_{-\boldsymbol{k}}^{(N)})^\dagger = \cosh(r)(\Hat{a}_{-\boldsymbol{k}}^{(0)})^\dagger + \sinh(r)e^{-i\varphi}\hat{a}^{(0)}_{\boldsymbol{k}},
    \label{eq:bogoliubov_1_v2}
\end{equation}
\begin{equation}
    \hat{a}^{(N)}_{\boldsymbol{k}} = \cosh(r)\hat{a}^{(0)}_{\boldsymbol{k}} + \sinh(r)e^{i\varphi}(\Hat{a}_{-\boldsymbol{k}}^{(0)})^\dagger.
    \label{eq:bogoliubov_2_v2}
\end{equation}
\end{subequations}

It is well-known that the action of a Bogoliubov transformation on the creation and annihilation operators is equivalent to that of a squeezing operator \cite{barnett2002methods}. Hence, by defining the complex squeezing parameter $\zeta = r\exp(i\varphi)$, we can introduce the following two mode squeezing operation:

\begin{equation}
    \hat{S}(\zeta) = \exp(\zeta\Hat{a}^{(0)}_{\boldsymbol{k}}\Hat{a}^{(0)}_{-\boldsymbol{k}}-\zeta^*(\Hat{a}_{\boldsymbol{k}}^{(0)})^\dagger(\Hat{a}_{-\boldsymbol{k}}^{(0)})^\dagger).
    \label{eq:squeezing}
\end{equation}

The squeezing operator in Eq. \ref{eq:squeezing} acts as the unitary time-evolution one in our theory and hence, it should be related to the system's Hamiltonian.\\
We recall, though, that in deriving the above formula, we have redefined the field amplitudes at two different times: the first one, when we did $B^{(N)}_{\boldsymbol{k}\sigma,a\pm} = \Tilde{B}^{(N)}_{\boldsymbol{k}\sigma,a\pm}\exp(\pm i\omega_k(NT+t_a)/n_a)$ and $B^{(N)}_{\boldsymbol{k}\sigma,b\pm} = \Tilde{B}^{(N)}_{\boldsymbol{k}\sigma,b\pm}\exp(\pm i\omega_k(NT+t_a)/n_b)$; this, we recall, enabled us to write the temporal boundary conditions in a $N$-independent manner and then calculate the $\boldsymbol{T}$-matrix. The second one was when we did $\hat{a}^{(N)}_{\boldsymbol{k}}\exp(i\theta_1)\longrightarrow\hat{a}^{(N)}_{\boldsymbol{k}}$ and $(\hat{a}^{(N)}_{-\boldsymbol{k}})^\dagger\exp(-i\theta_1)\longrightarrow(\hat{a}^{(N)}_{-\boldsymbol{k}})^\dagger$, which allowed us to finally introduce the squeezing operation in seen in Eq. \ref{eq:squeezing}.\\
The two transformations of the field amplitudes discussed above are actually equivalent to performing two diagonal unitary transformations. Since they are diagonal, they commute and can be combined to form a single diagonal unitary operation. This is equivalent to the non-linear term emerging from the discontinuity in the permittivity being the generator of the whole temporal evolution of the system and hence, to us working in an interaction-type picture, in which photon number-conserving terms do not contribute to the time dynamics (at least not explicitly).\\
Actually, the combined Hamiltonian for the $\pm\boldsymbol{k}$ modes is given by:

\begin{equation}
    \begin{split}
        \hat{H} & = \frac{\hbar\omega_k}{n_a}\left(\cosh^2(r) + \sinh^2(r)\right)\left((\hat{a}^{(0)}_{\boldsymbol{k}})^\dagger\hat{a}^{(0)}_{\boldsymbol{k}} + (\hat{a}^{(0)}_{-\boldsymbol{k}})^\dagger\hat{a}^{(0)}_{-\boldsymbol{k}} + 2\right) \\& +2\frac{\hbar\omega_k}{n_a}\cosh(r)\sinh(r)\left(e^{i\varphi}(\hat{a}^{(0)}_{\boldsymbol{k}})^\dagger(\hat{a}^{(0)}_{-\boldsymbol{k}})^\dagger + e^{-i\varphi}\hat{a}^{(0)}_{\boldsymbol{k}}\hat{a}^{(0)}_{-\boldsymbol{k}}\right)\\& = \hat{H}_1 + \hat{H}_2,
    \end{split}
\end{equation}

where $\hat{H}_{1,2}$ denote the photon number conserving and the non-linear parts of the Hamiltonian, respectively\footnote{We have also gone from a continuous spectrum to a discrete one; formally, this means the Dirac deltas need to be replaced by Kronecker ones.}. Thus, we see $\hat{H}_1$ can be eliminated by working in an interaction picture, in which only $\hat{H}_2$ is responsible for the temporal dynamics of the system.
\par
What follows is to calculate the matrix elements of the squeezing operator in the number state basis. To do so, we use a symmetrized version of equation \ref{eq:squeezing}, which can be found in \cite{barnett2002methods}:

\begin{equation}
    \hat{S}(\zeta) = \exp(-\tanh(r)e^{i\varphi}(\Hat{a}_{\boldsymbol{k}}^{(0)})^\dagger(\Hat{a}_{-\boldsymbol{k}}^{(0)})^\dagger)\exp(-\ln(\cosh(r))(\hat{n}^{(0)}_{\boldsymbol{k}}+\hat{n}^{(0)}_{-\boldsymbol{k}}+1))\exp(\tanh(r)e^{-i\varphi}\Hat{a}^{(0)}_{\boldsymbol{k}}\Hat{a}^{(0)}_{-\boldsymbol{k}}),
    \label{eq:squeezing_sym}
\end{equation}

Next, we calculate the state $\hat{S}(\zeta)\ket{n_{\boldsymbol{k}},m_{-\boldsymbol{k}}}$. If we expand in a power series the third term,

\begin{equation*}
\begin{split}
    &\exp(\tanh(r)e^{-i\varphi}\Hat{a}^{(0)}_{\boldsymbol{k}}\Hat{a}^{(0)}_{-\boldsymbol{k}})\ket{n_{\boldsymbol{k}},m_{-\boldsymbol{k}}} = \left(\sum_{l=0}^{\infty}\frac{1}{l!}(\tanh(r)e^{-i\varphi}\Hat{a}^{(0)}_{\boldsymbol{k}}\Hat{a}^{(0)}_{-\boldsymbol{k}}))^l\right)\ket{n_{\boldsymbol{k}},m_{-\boldsymbol{k}}}\\&=
    \sum_{l=0}^\infty\frac{(\tanh(r)e^{-i\varphi})^l}{l!}\sqrt{\frac{n!m!}{(n-l)!(m-l)!}}\Theta_{n-l}\Theta_{m-l}\ket{(n-l)_{\boldsymbol{k}},(m-l)_{-\boldsymbol{k}}},
\end{split} 
\end{equation*}

where $\Theta_{n-l}$ and $\Theta_{m-l}$ are Heaviside step functions, which give an upper bound on the summation index $l$, with $l\leq\min(m,n)$.\\
If we now act on the left with $\bra{n'_{\boldsymbol{k}},m'_{-\boldsymbol{k}}}$, since the first term in equation \ref{eq:squeezing_sym} is the adjoint of the third one, we have

\begin{equation*}
    \begin{split}
        &\bra{n'_{\boldsymbol{k}},m'_{-\boldsymbol{k}}}\exp(-\tanh(r)e^{i\varphi}(\Hat{a}_{\boldsymbol{k}}^{(0)})^\dagger(\Hat{a}_{-\boldsymbol{k}}^{(0)})^\dagger) = \\&
    \sum_{l'=0}^\infty\frac{(-\tanh(r)e^{i\varphi})^{l'}}{l'!}\sqrt{\frac{n'!m'!}{(n'-l')!(m'-l')!}}\Theta_{n'-l'}\Theta_{m'-l'}\bra{(n'-l')_{\boldsymbol{k}},(m'-l')_{-\boldsymbol{k}}} ,
    \end{split}
\end{equation*}

where the step functions yield the constraint $l'\leq\min(m',n')$.\\
Lastly, the middle term's action on the number states is trivial, with $\exp(-\ln(\cosh(r))(\hat{n}^{(0)}_{\boldsymbol{k}}+\hat{n}^{(0)}_{-\boldsymbol{k}}+1))\ket{j_{\boldsymbol{k}},j'_{-\boldsymbol{k}}} = \exp(-\ln(\cosh(r))(j+j'+1))\ket{j_{\boldsymbol{k}},j'_{-\boldsymbol{k}}} = (\cosh(r))^{-(j+j'+1)}\ket{j_{\boldsymbol{k}},j'_{-\boldsymbol{k}}}$. Hence, the matrix elements of the squeezing operator in the number state basis are given by:

\begin{equation*}
    \begin{split}
        &\bra{n'_{\boldsymbol{k}\lambda},m'_{-\boldsymbol{k}\lambda}}\hat{S}(\zeta)\ket{n_{\boldsymbol{k}\lambda},m_{-\boldsymbol{k}\lambda}}  = \sum_{l'=0}^{\min(m',n')}\frac{(-\tanh(r)e^{i\varphi})^{l'}}{l'!}\sqrt{\frac{n'!m'!}{(n'-l')!(m'-l')!}}\cdot\\&\sum_{l=0}^{\min(m,n)}\frac{(\tanh(r)e^{-i\varphi})^{l}}{l!}\sqrt{\frac{n!m!}{(n-l)!(m-l)!}}(\cosh(r))^{-(n+m-2l+1)}\delta_{n'-l',n-l}\delta_{m'-l',m-l}.
    \end{split}
\end{equation*}

From the two Kronecker deltas (which stem from the orthogonality between photon number states), we get the equations $n'-l'=n-l$ and $m'-l' = m-l$. If we take their difference, we obtain $n'-n = m'-m$, which is a statement of the conservation of momentum. Upon substitution of this constraint in the original equations, we obtain $l' = l + n'-n$, which allow us to eliminate the summation on the $l'$ index.\\
However, we recall that $l'\geq0\Longrightarrow l\geq n-n'$ and therefore, $l\geq\max(0,n-n')$. Likewise, $l'\leq\min(m',n')$, which yields $l\leq\min(m',n')+n-n'$.\\
We need to consider both $m'>n'$ and $m'<n'$ separately. In the former case ($m'>n'$), we get $l\leq n$, which is its natural bound, since the constraint $n'-n = m'-m$ means if $m'>n'$, then $m>n$; hence, $\min(m,n) = n$. On the other hand, for $m'<n'$, we get $l\leq n-(n'-m') = m$ and thus, we retain the same upper bound on $l$, since $m'<n'$ means $m<n$ and so, $\min(m,n) = m$. Taking all this into account, we finally arrive to   

\begin{equation}
\begin{split}
   &\bra{n_{\boldsymbol{k}}',m_{-\boldsymbol{k}}'}\hat{S}(\zeta)\ket{n_{\boldsymbol{k}},m_{-\boldsymbol{k}}} = 
   \frac{\left(-e^{i\varphi}\tanh(r)\right)^{n'-n}}{(\cosh(r))^{n+m+1}}\delta_{n'-n,m'-m}\\&\sum_{l = \max(0,n-n')}^{\min(n,m)}\frac{\sqrt{n'!m'!n!m!}}{l!(l+n'-n)!(n-l)!(m-l)!}(-\sinh^2(r))^l, 
\end{split}
    \label{eq:squeezing_matrix_elements}
\end{equation}
which is the equation found in the main text, save for the $C^l_{n,m;n',m'}$ defined in there.\\
So far we have been working with a plane wave expansion of the fields, but it is equally valid to use sines and cosines. In this case, the transfer matrix couples every mode to itself and upon quantization, the Bogoliubov transformation and the squeezing operator become one mode. The latter reads \cite{barnett2002methods,gerry2023introductory}:

\begin{equation}
    \hat{S}(\zeta) = \exp(\frac{1}{2}\left(\zeta(\hat{a}^{(0)}_{\boldsymbol{k}p})^2-\zeta^*((\hat{a}^{(0)}_{\boldsymbol{k}p})^\dagger)^2\right)),
    \label{eq:squeezing_sm}
\end{equation}

where $\boldsymbol{k}$ needs to be limited to half a sphere to avoid counting twice a given mode and where $p=c,s$ is an additional subscript which specifies the parity of the mode ($c$ for cosines and $s$ for sines). The polarization index has been dropped once again, but the parity one has been included to emphasize that the basis used in \ref{eq:squeezing} and \ref{eq:squeezing_sm} are not the same. However, the $T-$matrix does not couple modes whose parities differ and hence, the $p$ subscript could be equally omitted.\\
For the single mode squeezing operator there is a symmetrized formula similar to \ref{eq:squeezing_sym}, which reads \cite{barnett2002methods}:

\begin{equation}
    \hat{S}(\zeta) = \exp(-\frac{1}{2}\tanh(r)e^{i\varphi}((\hat{a}^{(0)}_{\boldsymbol{k}p})^\dagger)^2)\exp(-\ln(\cosh(r))\left(\hat{n}^{(0)}_{\boldsymbol{k}p}+\frac{1}{2}\right))\exp(\frac{1}{2}\tanh(r)e^{-i\varphi}(\hat{a}^{(0)}_{\boldsymbol{k}p})^2).
    \label{eq:squeezing_one_mode_sym}
\end{equation}

If we follow the same steps as we did for the two mode case, we arrive to

\begin{equation}
    \bra{n_{\boldsymbol{k}p}'}\hat{S}(\zeta)\ket{n_{\boldsymbol{k}p}} = \frac{\left(-\frac{1}{2}e^{i\varphi}\tanh(r)\right)^{\frac{n'-n}{2}}}{(\cosh(r))^{n+\frac{1}{2}}}\sum_{l = \max(0,\frac{n-n'}{2})}^{[\frac{n}{2}]}\frac{\sqrt{n'!n!}}{l!(l+\frac{n'-n}{2})!(n-2l)!}\left(-\frac{1}{4}\sinh^2(r)\right)^l,
    \label{eq:squeezing_one_mode_matrix_elements}
\end{equation}

where $[n/2]$ is the integer closest to $n/2$ (which needs not be even) and where $n'-n$ is a multiple of two, with the matrix elements being zero otherwise.\\
Lastly, we calculate for given squeezing strength $r$
the maximum photon pair creation probability $\text{Max(Prob)}$ for both the one and two mode squeezing cases.\\
In the latter, we assume we start with the state $\ket{n_{\boldsymbol{k}},0_{-\boldsymbol{k}}}$, where initially there are $n$ photons in the forward mode, with the backward one is its vacuum state. The transition probability towards $\ket{(n+m)_{\boldsymbol{k}},m_{-\boldsymbol{k}}}$ is then given by

\begin{equation}
    \text{Prob}(n,0;n+m,m) = \frac{(n+m)!}{n!m!}\frac{(\tanh(r))^{2m}}{(\cosh(r))^{2(n+1)}}.
\end{equation}

If we now differentiate with respect to $r$ and set the first derivative to zero, we arrive to the following maximum condition:

\begin{equation}
    \sinh^2(r) = m + \sqrt{m^2+4m(n+1)}.
\end{equation}

Substituting the above in the probability formula, we obtain an upper bound for the $\ket{n_{\boldsymbol{k}},0_{-\boldsymbol{k}}}\longrightarrow\ket{(n+m)_{\boldsymbol{k}},m_{-\boldsymbol{k}}}$ transition probability:

\begin{equation}
    \text{Max(Prob)}(n,0;n+m,m) = \frac{(n+m)!}{n!m!}\left(\frac{m + \sqrt{m^2+4m(n+1)}}{1+m + \sqrt{m^2+4m(n+1)}}\right)^m\frac{1}{\left(m + \sqrt{m^2+4m(n+1)}\right)^{n+1}},
\end{equation}

which decays monotonously with the difference $m-n$.\\
If we do the same for the one mode case for $n=0$, we get

\begin{equation}
    \text{Prob}(0;2m) = \frac{(2m)!}{4^m(m!)^2}\frac{(\tanh(r))^{2m}}{\cosh(r)},
\end{equation}

which has the extrema condition:

\begin{equation}
    \sinh^2(r) = m + \sqrt{m^2+2m}.
\end{equation}

\begin{figure*}
    \centering
    \vspace*{0cm}\hspace*{-0cm}\includegraphics[scale = 0.6]{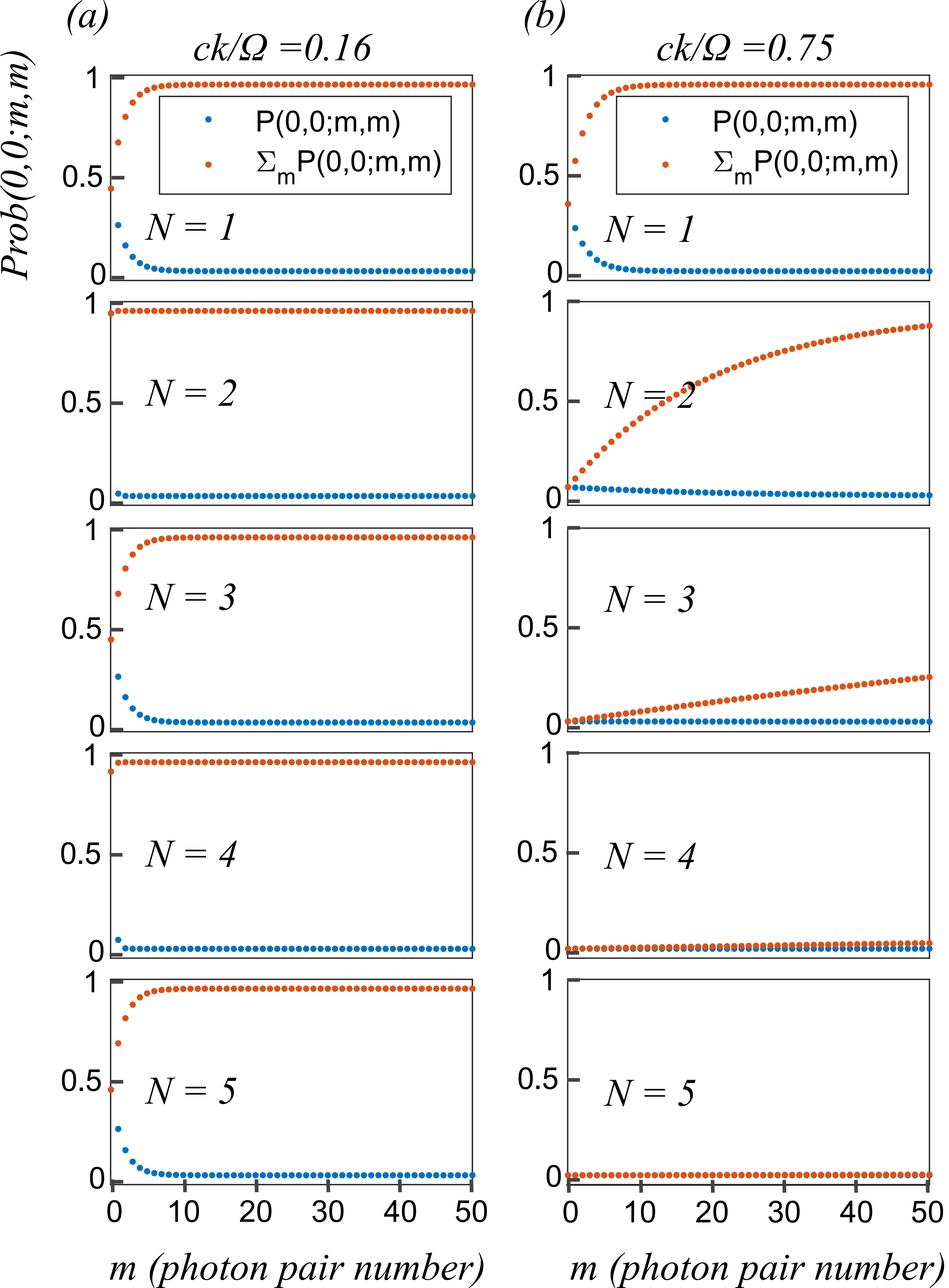}
    \caption{Photon pair creation probability $\text{Prob}(0,0;m,m)$ and the corresponding accumulated one, $\sum_{m'=0}^m\text{Prob}(0,0;m',m')$, for a mode within a band $(a)$, with real Floquet frequency, and another one within a bandgap (b), whose Floquet frequency is complex-valued. We consider $N=1,2,3,4\text{ and }5$ applications of the transfer matrix.}
    \label{fig:two_col_fig}
\end{figure*}

Upon substitution of the above in the probability formula, we arrive to

\begin{equation}
    \text{Max(Prob)}(0;2m) = \frac{(2m)!}{4^m(m!)^2}\left(\frac{m + \sqrt{m^2+2m}}{1+m + \sqrt{m^2+2m}}\right)^m\frac{1}{\sqrt{m + \sqrt{m^2+2m}}},
\end{equation}

which we present here for the sake of completeness.\\
In Fig. \ref{fig:two_col_fig} we plot $\text{Prob}(0,0;m,m)$ (blue dots) as well as $\sum_{m'=0}^m\text{Prob}(0,0;m',m')$ (red dots) for $N=1,2,3,4\text{ and }5$ periods of the PTC, and for a mode within a band, with $ck/\Omega=0.16$ (with real Floquet frequency), and another one within a momentum gap, with $ck/\Omega=0.75$ (with complex Floquet frequency). In the latter case, the probability for generating photons from the vacuum state gets squashed down as more periods are added to the PTC, with $\text{Prob}(0,0;m,m)$ becoming asymptotically uniform.

\section*{Photon number fluctuations and coherence}

In the present section, we calculate $\sigma(NT)$ and $g^{(2)}(NT)$ for number, coherent and thermal states, both one and two mode squeezed. We shall start with the latter, since the calculations are simpler.
\par
We recall that the photon number variance to average ratio was given by

\begin{equation}
    \sigma(NT) = \frac{\langle(\Delta\Hat{n}^{(N)}_{\boldsymbol{k}})^2\rangle_0}{\langle\Hat{n}^{(N)}_{\boldsymbol{k}}\rangle_0}.
\end{equation}

Thus, we need to calculate both $\langle\hat{n}^{(N)}_{\boldsymbol{k}}\rangle_0$ and $\langle(\hat{n}^{(N)}_{\boldsymbol{k}})^2\rangle_0$. We assume the backward mode is initially in its vacuum state, thus considering the states $\ket{n_{\boldsymbol{k}},0_{-\boldsymbol{k}}}$, $\ket{\beta_{\boldsymbol{k}},0_{-\boldsymbol{k}}}$ and $\exp(-\hbar\omega_k\hat{n}^{(0)}_{\boldsymbol{k}}/k_BT)/Z\otimes\ket{0_{-\boldsymbol{k}}}\bra{0_{-\boldsymbol{k}}}$.\\
For $\ket{\psi_0} = \ket{n_{\boldsymbol{k}},0_{-\boldsymbol{k}}}$, the average of the photon number and its square read

\begin{subequations}
    \begin{equation}
    \langle\hat{n}^{(N)}_{\boldsymbol{k}}\rangle_0 = \cosh^2(r)n + \sinh^2(r)
    \end{equation}
    and
    \begin{equation}
        \langle(\hat{n}^{(N)}_{\boldsymbol{k}})^2\rangle_0 = \cosh^4(r)n^2+3\cosh^2(r)\sinh^2(r)n + \cosh^2(r)\sinh^2(r) + \sinh^4(r).
    \end{equation}
\end{subequations}

For the coherent state $\ket{\psi_0} = \ket{\beta_{\boldsymbol{k}},0_{-\boldsymbol{k}}}$, we have

\begin{subequations}
    \begin{equation}
    \langle\hat{n}^{(N)}_{\boldsymbol{k}}\rangle_0 = \cosh^2(r)|\beta|^2 + \sinh^2(r)
    \end{equation}
    and
    \begin{equation}
        \langle(\hat{n}^{(N)}_{\boldsymbol{k}})^2\rangle_0 = \cosh^4(r)|\beta|^4+\left(\cosh^4(r) + 3\cosh^2(r)\sinh^2(r)\right)|\beta|^2 + \cosh^2(r)\sinh^2(r) + \sinh^4(r).
    \end{equation}
\end{subequations}

Finally, the thermal field $\rho_0 = \exp(-\hbar\omega_k\hat{n}^{(0)}_{\boldsymbol{k}}/k_BT)/Z\otimes\ket{0_{-\boldsymbol{k}}}\bra{0_{-\boldsymbol{k}}}$ has

\begin{subequations}
    \begin{equation}
        \langle\hat{n}^{(N)}_{\boldsymbol{k}}\rangle_0 = \cosh^2(r)\bar{n} + \sinh^2(r)
    \end{equation}
    and
    \begin{equation}
        \langle(\hat{n}^{(N)}_{\boldsymbol{k}})^2\rangle_0 = 2\cosh^4(r)\bar{n}^2+\left(\cosh^4(r) + 3\cosh^2(r)\sinh^2(r)\right)n + \cosh^2(r)\sinh^2(r) + \sinh^4(r),
    \end{equation}
\end{subequations}

where $\bar{n}$ is the average photon number as given by the Maxwell-Boltzmann thermal distribution \cite{gerry2023introductory}.\\
If one now calculates the variance to average photon number explicitly, it can be seen $\sigma(NT)/\sigma(0)\geq1$,  with the equality corresponding to $r = 0$ and the temporal modulation being transparent to the mode. Therefore, fluctuations in the energy can only be enhanced or remain the same, but never be reduced.
\par
The second order coherence function is defined following Glauber's seminal paper \cite{glauber1963quantum} for the case of zero time delay, as seen in \cite{gerry2023introductory}:

\begin{equation}
    g^{(2)}(NT) = \frac{\langle(\Hat{n}^{(N)}_{\boldsymbol{k}})^2\rangle_0-\langle\Hat{n}^{(N)}_{\boldsymbol{k}}\rangle_0}{(\langle\Hat{n}^{(N)}_{\boldsymbol{k}}\rangle_0)^2},
\end{equation}
where the $NT$ in the argument of $g^{(2)}$ is not to be confused with the $\tau$ usually found in the definition of the time-delayed second order coherence function: here we calculate the instantaneous $g^{(2)}$ (zero time-delay then) of the squeezed quantum field at a time $t = NT$.\\
Making use of all the formulae derived for $\langle\hat{n}^{(N)}_{\boldsymbol{k}}\rangle_0$ and $\langle(\hat{n}^{(N)}_{\boldsymbol{k}})^2\rangle_0$, we can calculate $g^{(2)}(NT)$ rather easily for all three states considered. If one does so, it can be seen that $g^{(2)}(NT)=2$ for the forward thermal field/backward vacuum, and that the $r\longrightarrow\infty$ limit yields $g^{(2)}\longrightarrow1+1/(n+1)$ and $g^{(2)}\longrightarrow1+1/(|\beta|^2+1)+|\beta|^2/(|\beta|^2+1)^2$ for forward number and forward coherent states.
\par
Next, we calculate the fluctuations and the second order coherence function for the one mode squeezing case for the sake of completeness.\\
For a one-mode squeezed number state, we have

\begin{subequations}
    \begin{equation}
        \langle\Hat{n}^{(N)}_{\boldsymbol{k}p}\rangle_0 = \left(\cosh^2(r) + \sinh^2(r)\right)n + \sinh^2(r)
    \end{equation}

    and
\begin{equation}
 \begin{split}
     &\langle(\Hat{n}^{(N)}_{\boldsymbol{k}p})^2\rangle_0 = \left(\cosh^4(r) + \sinh^4(r) + 4\cosh^2(r)\sinh^2(r)\right)n^2 + \\&2\left(\sinh^4(r)+2\cosh^2(r)\sinh^2(r)\right)n + \sinh^4(r) + 2\cosh^2(r)\sinh^2(r),
 \end{split}
\end{equation}
    
\end{subequations}

while for the coherent one it is
\begin{figure*}
    \centering
    \vspace*{0cm}\hspace*{-0cm}\includegraphics[scale = 0.5]{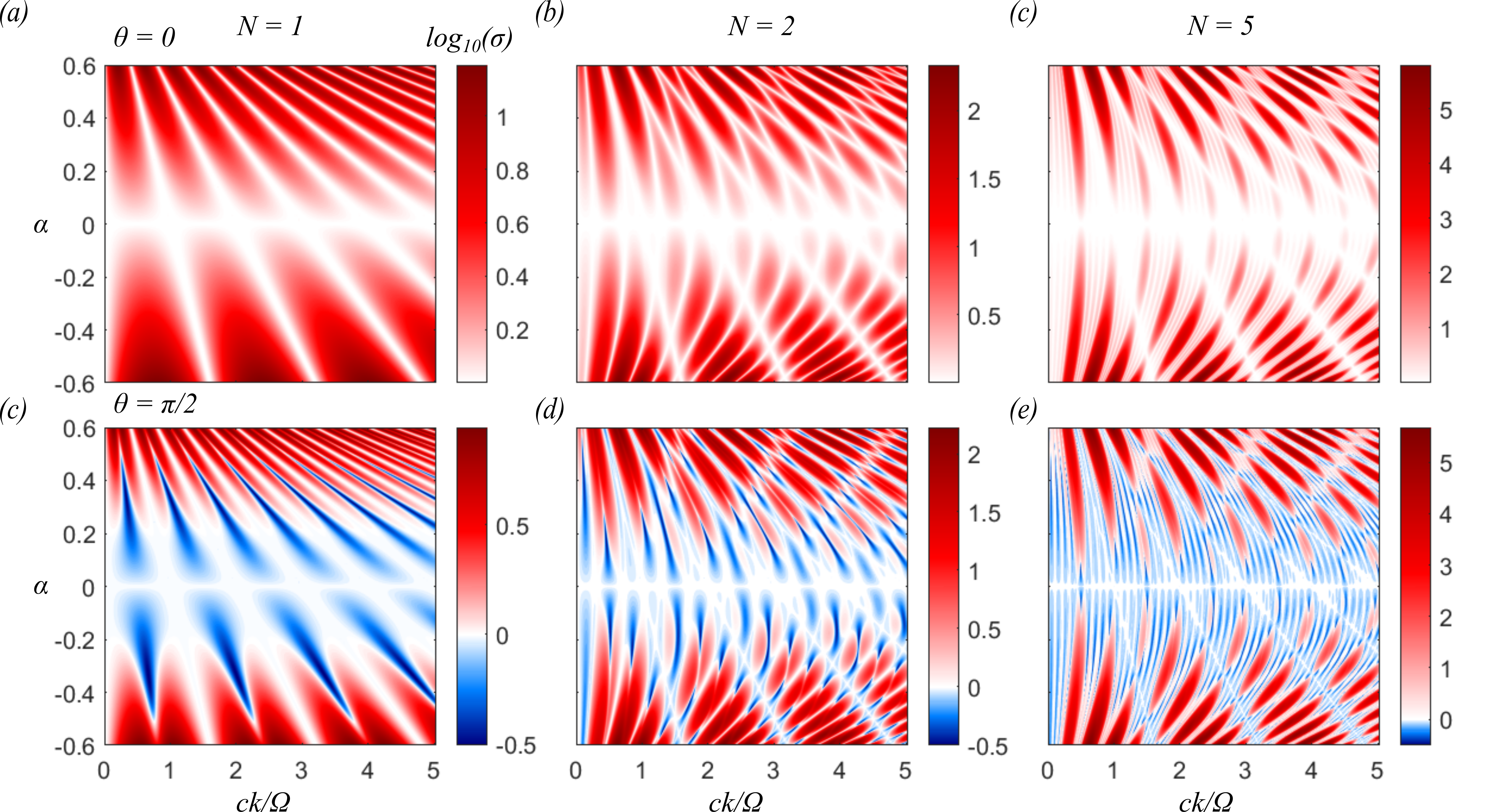}
    \caption{Fluctuations of a one one-mode squeezed coherent state for $N = 1,2 \text{ and } 5$ applications of the transfer matrix, for values of the phase of $\theta=0 \text{ and }\theta = \pi/2$ and a fixed amplitude of $|\beta| = 10$. Blue coloured regions correspond to fluctuations smaller than unity and hence, sub-Poissonian.}
    \label{fig:wide_fig_1}
\end{figure*}

\begin{subequations}
    \begin{equation}
        \langle\Hat{n}^{(N)}_{\boldsymbol{k}p}\rangle_0  = \Big(\cosh^2(r) + \sinh^2(r) + 2\cosh(r)\sinh(r)\cos(\varphi-2\theta)\Big)\abs{\beta}^2 + \sinh^2(r)
    \end{equation}
    and
    \begin{equation}
        \begin{split}
            &\langle(\Hat{n}^{(N)}_{\boldsymbol{k}p})^2\rangle_0  = \Big(\cosh^4(r) + \sinh^4(r) + 4\cosh^2(r)\sinh^2(r) + \\&4\left((\cosh(r)\sinh^3(r)+\cosh^3(r)\sinh(r)\right)\cos(\varphi-2\theta)+2\cosh^2(r)\sinh^2(r)\cos\left(2(\varphi-2\theta)\right)\Big)|\beta|^4+\\&
            \Big(\cosh^4(r) + 3\sinh^4(r) + 8\cosh^2(r)\sinh^2(r)+4\left(\cosh^3(r)\sinh(r)+2\cosh(r)\sinh^3(r)\right)\cos(\varphi-2\theta)\Big)|\beta|^2\\&
            + \sinh^4(r) + 2\cosh^2(r)\sinh^2(r)
        \end{split}
    \end{equation}
\end{subequations}

where $\beta = |\beta|\exp(i\theta)$.\\
Lastly, for a one-mode squeezed thermal state, we get

\begin{subequations}
    \begin{equation}
        \langle\Hat{n}^{(N)}_{\boldsymbol{k}p}\rangle_0 = \left(\cosh^2(r) + \sinh^2(r)\right)\Bar{n} + \sinh^2(r)
    \end{equation}
    and
 \begin{equation}
 \begin{split}
     &\langle(\Hat{n}^{(N)}_{\boldsymbol{k}p})^2\rangle_0 = 2\left(\cosh^4(r) + \sinh^4(r) + 4\cosh^2(r)\sinh^2(r)\right)\bar{n}^2 + \\&\left(\cosh^4(r) + 3\sinh^4(r)+8\cosh^2(r)\sinh^2(r)\right)\bar{n} + \sinh^4(r) + 2\cosh^2(r)\sinh^2(r).
 \end{split}
\end{equation}
    
\end{subequations}

It can be proved mathematically that once again, the fluctuations for number and thermal one-mode squeezed states can only grow or remain the same, as in the two mode case. However, this is not the case for coherent ones, in which fluctuations are phase sensitive and can actually become sub-Poissonian for small but not zero reflectivity (see figure \ref{fig:wide_fig_1}). The phase-sensitivity of the one mode squeezing is reminiscent of that of the degenerate parametric amplifier \cite{boyd2008nonlinear,drummond2014quantum}, in which the idler and signal modes share the same frequency. More information about the coherence and statistics of one mode squeezed quantum fields can be found in \cite{kim1989properties}. The main difference lies in the fact that a one-mode squeezed thermal state has its $g^{(2)}$ increased and hence, more than two photons are detected on average by a suitable photodetector.
\bibliographystyle{unsrt}
\bibliography{sm}